\documentclass[11pt,a4paper]{article}

\usepackage{color}
\usepackage{amssymb}
\usepackage{amsmath}

\def\na{\nu_1}
\def\nb{\nu_2}
\def\nc{\nu_3}

\def\z{{\zeta}}

\def\mmu{\delta}
\def\a{m_1}
\def\b{m_2}
\def\c{m_3}
\def\Z{Z}
\def\bC{{\bf C}}
\def\bR{{\bf R}}

\def\bZ{{\bf Z}}
\def\Im{\mathop{\rm Im}\nolimits}
\def\Re{\mathop{\rm Re}\nolimits}

\def\tg{\mathop{\rm tg}\nolimits}
\def\th{\mathop{\rm th}\nolimits}
\def\ch{\mathop{\rm ch}\nolimits}
\def\sh{\mathop{\rm sh}\nolimits}

\def\TT{{\cal T}}

\def\wh{\widehat}
\def\vsk{{\vskip 4mm }}

\def\half{{\scriptstyle{1 \over 2}}}
\def\interior#1{\setbox1=\hbox{$#1$}\rlap{$#1$}\kern0.4\wd1\raise1.1\ht1%
\hbox{$\scriptstyle \circ$}}

\def\boxit#1#2{\setbox1=\hbox{\kern#1{#2}\kern#1}%
\dimen1=\ht1 \advance \dimen1 by #1 \dimen2=\dp1 \advance \dimen2 by #1
\setbox1=\hbox{\vrule height\dimen1 depth\dimen2\box1\vrule}%
\setbox1=\vbox{\hrule\box1\hrule}%
\advance \dimen1 by .4pt \ht1=\dimen1 \advance \dimen2 by .4pt \dp1=\dimen2
\box1\relax}
\def\endprf{\raise .5ex\hbox{\boxit{2pt}{\ }}}

\def\ifundefined#1{\expandafter\ifx\csname#1\endcsname\relax}

\def\beq{\begin{equation}}
\def\endq{\end{equation}}
\def\beqa{\begin{eqnarray}}
\def\endqa{\end{eqnarray}}

\newcommand\n{\kappa}

\newcommand{\sd}{{ S}_{d-1}}

\let\UnmodifSec=\section
\renewcommand{\section}{\setcounter{equation}{0}\UnmodifSec}

\newtheorem{remark}{Remark}[section]

\newtheorem{formula}{Formula}

\def\bbeta{{\rho}}
\def\aalpha{{\mu}}

\def\formdif{{\alpha}}
\def\z{{\zeta}}
\def\bC{{\bf C}}
\def\bR{{\bf R}}

\def\bZ{{\bf Z}}

\def\Im{\mathop{\rm Im}\nolimits}
\def\Re{\mathop{\rm Re}\nolimits}

\def\tg{\mathop{\rm tg}\nolimits}
\def\th{\mathop{\rm th}\nolimits}
\def\ch{\mathop{\rm ch}\nolimits}
\def\sh{\mathop{\rm sh}\nolimits}

\def\TT{{\cal T}}

\def\wh{\widehat}

\def\half{{\scriptstyle{1 \over 2}}}
\def\interior#1{\setbox1=\hbox{$#1$}\rlap{$#1$}\kern0.4\wd1\raise1.1\ht1%
\hbox{$\scriptstyle \circ$}}

\def\boxit#1#2{\setbox1=\hbox{\kern#1{#2}\kern#1}%
\dimen1=\ht1 \advance \dimen1 by #1 \dimen2=\dp1 \advance \dimen2 by #1
\setbox1=\hbox{\vrule height\dimen1 depth\dimen2\box1\vrule}%
\setbox1=\vbox{\hrule\box1\hrule}%
\advance \dimen1 by .4pt \ht1=\dimen1 \advance \dimen2 by .4pt \dp1=\dimen2
\box1\relax}
\def\endprf{\raise .5ex\hbox{\boxit{2pt}{\ }}}

\def\ifundefined#1{\expandafter\ifx\csname#1\endcsname\relax}

\def\beq{\begin{equation}}
\def\endq{\end{equation}}
\def\beqa{\begin{eqnarray}}
\def\endqa{\end{eqnarray}}

\def\P{{\bf P}}
\def\Q{{\bf Q}}

\renewcommand{\cosh}{\ch}\renewcommand{\sinh}{\sh}
\renewcommand{\tanh}{\th}\renewcommand{\tan}{\tg}
\usepackage{graphicx}  
\def\funame{\Phi}

\begin{document}

\title{Loops in de Sitter space\footnote{To Jacques Bros and Michel Gaudin. In loving memory}}
\author{Sergio L. Cacciatori$^{(a,b)}$, Henri Epstein$^{(c)}$ and Ugo Moschella$^{(a,b)}$\\ \\ 
$^{(a)}$Disat, Universit\`a dell'Insubria, Como \\  $^{(b)}$INFN, Sezione di Milano, Italia\\ $^{(c)}$IHES, Bures-sur-Yvette, France\\}
\maketitle
\abstract{We discuss general one and two-loops banana diagrams with arbitrary masses on the de Sitter spacetime by using direct methods of dS quantum field theory in the dimensional regularization approach. In the one-loop case we also compute the effective potential for an $O(N)$ model in $d=4$ dimension as an explicit function of the cosmological constant $\Lambda$, both exactly and perturbatively up to order $\Lambda$. For the two-loop case we show that the calculation is made easy thanks to a remarkable  K\"all\'en-Lehmann formula that has been in the literature for a while. We discuss the  divergent cases at $d=3$ using a contiguity formula for generalized hypergeometric functions and we extract the dominant term at $d=4$ proving a general formula to deal with a divergent hypergeometric series.}
\section{Introduction}
General Relativity and the Standard Model of Particle Physics are the greatest successes of theoretical physics and are among the highest  achievements of mankind.
Starting from the researches of the pre-Socratic philosophers of the Greek world of the VI century B.C., it took almost three millennia  to reach this level of comprehension of Nature \cite{um}. 

On its side, General Relativity offers the natural framework for gravity and cosmology, not only by providing equations that plausibly describe the evolution of the universe and of the objects that populate it, but also as a source of  ideas and technical tools  for improving and sometimes changing the way we  observe and measure the sky; 
examples are the already mature gravitational lensing and microlensing techniques 
and the brand new multi-messenger astronomy. 
On the other side, the Standard Model gives a deep understanding of the microscopic world and there is no need to recall its great successes here. 

The mysterious fact remains that all attempts  to construct a single coherent theory that includes both of them have consistently failed.
Nevertheless, cosmology, astrophysics and the physics of elementary particles are nowadays inextricably interconnected and even an incomplete understanding of the history and  the dynamics of the universe requires some form of coexistence of the two theories.  At the moment, the best thing that is available is Quantum Field Theory (QFT) on a  curved background, possibly including a quantum linearized gravitational field and backreaction. 

QFT in an inertial frame of flat Minkowski space is quite well understood, mainly at the  perturbative level but  non-perturbative methods have also  been developed over the years. The necessity of improving our computational skills for a deeper understanding of the  Standard Model has recently generated a great flurry of activity;  new strategies for computing Feynman-like integrals in  QFT  and  GR (especially for gravitational waves) are abundant in contemporary literature. 
Here are a few examples:

-- the proposed reformulation of perturbative QFT in terms of positive Grassmannian geometry in a complexified momentum space, leading to the notion of Amplituhedron and its generalizations \cite{Arkani1}--\cite{Arkani6}; 

-- the traditional method of integration by parts to uncover relations among different Feynman integrals in order to reduce their calculation to a subset of Master Integrals; the system of differential equations they have to satisfy \cite{Chetyrkin}--\cite{Laporta2000}; 

-- the cohomological techniques based on the interpretation of Feynman integrals as periods of generators of a suitable twisted-cohomology \cite{Mastrolia2019}--\cite{Chen2022}; 

 -- viewing Feynman integrals as a linear space with  the intersection product of the twisted cohomology used scalar product to determine bases of Master Integrals, Picard Fuchs-equations and so on \cite{Giroux2023}--\cite{Brunello:2023fef};
 
 -- other methods that consist in investigating the (algebraic) geometry underlying Feynman integrals  relating them to suitable Calabi-Yau varieties   \cite{Bonisch}--\cite{Frellesvig:2023bbf}. 
\vskip 10 pt
The crucial common ingredient of all the above approaches is the Fourier momentum-space  representation of Feynman integrals. Unfortunately, such representation is not available on curved backgrounds because  translation invariance, which is the foundation of Fourier analysis, pertains only to flat space\footnote{Even in flat space, a non trivial topology may limit the effectiveness of the momentum space formulation.};
the main road for calculating amplitudes in curved spacetimes is  to do it in position space. 
This approach is surprisingly efficient also in Minkowski space: in  \cite{Cacciatori2023a} we presented a study of one and two-loop diagrams in position space, improving by a margin the existing literature and also providing new insights into the method of partial integration.

In this paper we perform the same loop calculations in the de Sitter universe working in  position space at Euclidean times. 
The status of the de Sitter spacetime is however rather exceptional. Even in the absence of a linear momentum space, a complete harmonic analysis has been available for some time for de Sitter quantum field theory \cite{bgm,bm,bemgen,Bros} but this possibility has  not yet been fully exploited in applications (see however \cite{us}). 
We intend to fill this gap and show how this bunch of  rigorous methods and results is also  very effective to compute exact expressions for loop integrals which may in turn prove to be relevant for cosmology. 

In particular, we consider one and two-loop  diagrams  with no external legs; calculations are performed without relying on particular choices of coordinates on the de Sitter manifold;  the results are explicit  exact  formulae for loop integrals with two and, respectively, three independent scalar fields;  three different masses enter in the loop, no conformal invariance is supposed, no room for bootstrapping anything.

Our results generalize to the de Sitter spacetime the state of the art of the knowledge available in  flat space \cite{Cacciatori2023a,FordJones}. 
We expect that the methods and the tools exposed in this papers will be useful to face other calculations involving
loops on the (complex) de Sitter manifold.

The paper is organized as follows:
in Sec. \ref{sec1}, after recalling some generalities about de Sitter QFT, we write multiloop (banana) diagrams with no external legs in a form suitable for their computation in the rest of the paper.
Sec. \ref{1loop} is devoted to the one-loop diagram  with two independent masses. 

The proof of the  main formula (\ref{finalres}) is performed in position space and crucially relies on the identification of the two-point function as a Legendre function of the first kind (a Ferrers function of the first kind for the Euclidean propagator) as opposed to an equivalent but otherwise  less specific hypergeometric function $_2F_1$. 

 In Sec. \ref{sec2} we discuss an application: we compute the one-loop effective potential for the $O(N)$ model on the  de Sitter manifold in dimension $d=4$, exact at one-loop. Computations of effective potentials on the de Sitter background have sporadically appeared in the literature \cite{Esposito}--\cite{Serreau:2011fu} but there exists to date no rigorous deduction based on a formulation sufficiently general to allow for systematic strategies as is the case in flat spacetime. 

In Sec. \ref{flat} we compute the flat limit of the effective potential which exactly reproduces the well-known  result in Minkowski space. It is worthwhile  to underline already here that this is indeed a non-trivial fact, as the flat limit relates quantities that are the outcome of integrals of different functions over different manifolds and is not just the asymptotic value of some functions close to a given event. We express the exact one-loop potential as a function of the cosmological constant and show how to compute it perturbatively at any desired order, when the cosmological constant is small. 

In Sec. \ref{2loop}  we focus on the two-loop diagram with no external legs with three arbitrary masses; we study the  diagram in dimensional regularization for an arbitrary complex dimension of the de Sitter manifold and produce exact formulae for it in Eqs. (\ref{general}), (\ref{ad}), (\ref{i1bis}), (\ref{bd}) and (\ref{ii2}). The crucial ingredient allowing for a full solution of the problem is the K\"all\'en-Lehmann representation of the product of two propagators that was explicitly constructed a few years ago \cite{us}.

In the remaining sections we use our explicit formulae to study the diagram at integer spacetime dimension. We start with a discussion of odd negative spacetime-dimensions which also provides a check for our formulae; then we   produce a detailed study  of the  non-trivial cases $d=2$ and $d=3$. 

Extracting the finite part in $d=4$  can be done using the contiguity relations discussed in Appendix (\ref{pole})  precisely as it is  done for the $d=3$ case in Appendix \ref{d=3}. In $d=4$ the situation is complicated by the presence of a double pole; formulae for the finite part of the diagrams are too long to be reproduced here. We limit the  discussion to the residues that we  compute by using the Erd\'elyi-Tricomi theorem.

\section{Banana  integrals on the Euclidean  sphere} 
\label{sec1}
\subsection*{Geometry}
The easiest way to look at either the real or the complex $d$-dimensional de Sitter manifolds is to visualize them 
as subsets of the complex Minkowski spacetime with one 
spacelike dimension more. This viewpoint allows for a natural 
description of the fundamental tubular domains encoding the spectral 
condition of de Sitter quantum field theory \cite{bgm,bm,bemgen}.
  
Let therefore $M_{d+1}$
be the real
$(d+1)$-dimensional Minkowski space-time and $M_{d+1}^{(c)}$
be its complexification. 
In a chosen Lorentz frame 
the scalar product of two (complex) events is  
\begin{equation} z_1\cdot z_2  = z^0_1 z^0_2-z^1_1 z^1_2 - \ldots -z^d_1 z^d_2.
\end{equation}
 The future cone $V_+$ and the future and past
tubes $T_\pm$ of the (complex) Minkowski spacetime are defined as follows:
\begin{eqnarray}
V_+ &=& \{x\in M_{d+1}\ :\ x\cdot x >0,\ \ \ x^0> 0\},\\
T_\pm &=& \{x+iy \in M_{d+1}^{(c)}\ :\ y\in \pm V_+\}\ .
\label{s.1.2}\end{eqnarray}
The tubes $T_\pm$ are the geometrical sets corresponding to the spectral condition which requires the positivity of the spectrum of the energy operator in every Lorentz frame \cite{Streater}. This is the very characteristic property at the heart of QFT at zero temperature. All the well-known features of QFT, and above all the Euclidean formulation and renormalization, depend on it.  
 
 The real de Sitter  universe  may be represented as  the  one-sheeted hyperboloid immersed in $M_{d+1}$:
\beq
dS_d = \{x \in M_{d+1}\ :\ x\cdot x = -R^2=-1\}\ ;
\label{s.2}\endq
the same definition, {\em mutatis mutandis},  holds for its complexification:
\beq
dS_d^{(c)} = \{z \in M_{d+1}^{(c)}\ :\ z\cdot z = -R^2=-1\}\ .
\label{s.2}\endq
The de Sitter invariant  complex variable $\zeta$ is the scalar product in the ambient spacetime of  two complex events $z_1, z_2 \in dS_d^{(c)}$:  
\begin{equation}
    \zeta= z_1\cdot z_2.
\end{equation} 
Two real events $x_1$ and $x_2$ in $dS_d$ are timelike separated if and only if 
\begin{equation}
(x_1-x_2)^2  = - 2 - 2 x_1\cdot x_2 >0. \label{causal}
\end{equation}
The future and past tuboids $\TT_\pm$ are the intersections of the ambient tubes $T_\pm$ with the complex de Sitter manifold:
\beq
\TT_\pm = \{x+iy \in X_d^{(c)}\ :\ y\in \pm V_+\}\ .
\label{s.2.1}\endq

\subsection*{Harmonic analysis}
A natural  basis of
plane-wave solutions of the de Sitter Klein-Gordon equation
\begin{equation} \Box \psi (z)   +m^2 \psi (z) =0,\label{KGnu}\end{equation}
is 
parameterized by the choice of a lightlike vector $\xi\in C^+=\partial V^+$ and a complex number $\lambda$ which is in turn parametrized by the spacetime dimension and another complex number $\nu$ as follows:
\begin{equation}
\psi_\lambda(z,\xi) = \left({ z\cdot \xi}\right)^{\lambda}  \label{waves}, \ \ \ \ \lambda=-\frac{d-1}2 + i \nu.
\end{equation}
The parameters $\lambda$ and $\nu$ are related to the complex mass squared:
\begin{equation}
m^2 = -\lambda(\lambda+d-1) = \frac {(d-1)^2}{4} +\nu^2. \label{cmass}
\end{equation}
Plane waves are well-defined and analytic in each of the tubes
$\cal T^+$ and $\cal T^-$ \cite{bgm,bm}. 
Of course their squared mass is real and positive only when: 

\noindent a) $\nu$  is real; this correspond in a group-theoretical language to the principal series of unitary representations of the Lorentz group; 

\noindent b)  $\nu$ is purely imaginary such the $|\nu|<\frac{d-1}{2} $; this correspond to the complementary series of unitary representations of the Lorentz group.
\vskip 10 pt

In de Sitter spacetime there is no global timelike Killing vector. A spectral condition may however  be formulated as the following requirement of {\em normal analyticity} \cite{bm}:  the two-point distributions are  boundary values of functions analytic in the  domain ${\cal T}_-\times {\cal T_+}$; this condition, together with de Sitter invariance and the Canonical Commutation Relations,   selects a unique two-point function\footnote{It is called - for a strange historical habit -  Bunch-Davis vacuum. Actually, it was W. Thirring the first to find it.} for any de Sitter Klein-Gordon field \cite{bm}:

\vskip 10 pt

\noindent \underline {\bf Main result}:   \cite{bgm,bm}
{\em the normally analytic canonical  Wightman function of a de Sitter Klein-Gordon field in spacetime dimension $d$ whose complex mass squared 
$m^2 =  \frac{(d-1)^2}{4} +\nu^2$ is parametrized by a complex parameter $\nu$, has the following spectral representation in  plane waves and is holomorphic  for $z_1\in {\cal T}_-$ and  $z_2\in {\cal T}_+$:}
\begin{align}
W^d_\nu (z_1,z_2)   &= \frac{\Gamma\left(\frac{d-1}2 + i \nu\right)\Gamma\left(\frac{d-1}2 - i \nu\right)
e^{\pi\nu}}{2^{d+1}\pi^d}
\int_{\gamma}(\xi\cdot z_1)^{-\frac{d-1}2 - i \nu}(\xi\cdot z_2)^{-\frac{d-1}2 + i \nu}\,\formdif(\xi)  \label{uup}
\\ = w^d_\nu(\zeta)  &= \frac {\Gamma\left(\frac{d-1}2 + i \nu\right)\Gamma\left(\frac{d-1}2 - i \nu\right)}{2 (2\pi)^{d/2}}
(\zeta^2-1)^{-\frac {d-2}4}\,P_{ -{\frac 12}+ i \nu }^{-\frac {d-2}2}(\zeta) \label{leg2p}
 \\  &= 
\frac {\Gamma\left(\frac{d-1}2 + i \nu\right)\Gamma\left(\frac{d-1}2 - i \nu\right)}{(4\pi)^{d/2}\Gamma\left(\frac d2\right)}
{}_2F_1\left(\frac{d-1}2 + i \nu,\frac{d-1}2 - i \nu;\ {\frac d2};\ \frac {1-\zeta}2\right ).
\label{wig}
\end{align}
In  standard coordinates, the ($d-1$)-form $\alpha(\xi)$  in Eq. (\ref{uup}) is written 
\begin{equation} \alpha(\xi) = (\xi^0)^{-1} \sum_{j=1}^d (-1)^{j+1}
\xi^j\,d\xi^1\ldots\ \wh{d\xi^j}\ldots\ d\xi^d\ .
\label{f.16}\end{equation}
$\gamma$
denotes any
($d-1$)-cycle in the forward
light-cone $C^+$.  (\ref{uup}) does not depend on the choice of $\gamma$ being the integral of a closed differential form. In particular we may choose the  unit sphere
$\sd$ (equipped with its canonical orientation):
\begin{equation}
 \gamma_0=\sd= C^+ \cap \{\xi\ :\ \xi^0 =1\}
=  \{ \xi \in C^+ : {\xi^1}^2 + \ldots + {\xi^d}^2 =
1\}.
\end{equation}
With this choice $\alpha(\xi)$  coincides
with the rotation invariant measure $d\xi$ on  $\sd$ normalized as
follows:
\begin{equation}
\omega_{d}=\int_{\gamma_0}d\xi  = \frac{2\pi^\frac d2}{\Gamma\left(\frac d2\right)}.
\label{norms}
\end{equation}
By chosing any two points in the respective tubes one can show the validity of Eq. (\ref{leg2p}). A crucial fact is that the Legendre function \cite{bateman} of the first kind in that  formula  
\beq
P_\nu^\mu(\zeta) = {1\over \Gamma(1-\mu)}
\left({\zeta+1\over \zeta-1} \right )^{\mu\over 2}
F \left (-\nu,\ 1+\nu\ ;\ 1-\mu\ ;\ {1-\zeta\over 2}\right )\ 
\label{r.50}\endq
is  holomorphic in the cut-plane $\bC\setminus (-\infty,\ 1]$
but the reduced two-point function  
$w^{d} _\nu(\zeta)$ 
is holomorphic in the  larger domain\footnote{The prefactor $(\zeta^2-1)^{-{d-2 \over 4}}$ exactly compensates the singularity of  $P_{ -{1\over 2}+ i \nu }^{-{d-2 \over 2}}(\zeta) $ at $\zeta=1$, making the reduced two-point function regular there.}  
\beq \zeta  = z_1\cdot z_2 \in \Delta = \{\bC\setminus (-\infty,\ -1]\},     
\label{r.1.2}\endq 
i.e. everywhere except on the causality cut (\ref{causal}); this is the {\em maximal analyticity property}. 

When $\nu$  is either real or  is purely imaginary and such that  $|\nu|<\frac{d-1}{2} $  the corresponding two-point function (\ref{uup}) is {\em positive-definite} and admits a direct quantum probabilistic interpretation. 

When $\frac{d-1}2 + i \nu = -n$, where $n$ is zero or a positive integer, a more involved but yet  acceptable quantum interpretation of the above formula is also possible; on the de Sitter universe there exist {\em  tachyonic fields} having no counterpart in flat space \cite{bemt,tachyons}.

Finally, the Schwinger function (in short: the propagator) is the restriction of the maximally analytic two-point function to the Euclidean sphere. It can be obtained as follows:  in Eq. (\ref{uup}) choose the two points as follows 
\begin{align}
& z_1=\left(\begin{array}{c}
\sinh(-i\epsilon)  \\
0 \\
\vdots \\
0\\
\cosh(-i\epsilon) \\
\end{array}
\right),
 \ \ \ \ \ z_2(s)= \left(
\begin{array}{c}
\sinh(i s) 
\\
0\\
\vdots\\
0\\
\cosh( is) 
\end{array}
\right), \ \ \ 0<s<\pi,
\end{align}
so that $  z_1\cdot 
z_2(s) = -\cos(s-i\epsilon )$; we get 
 the following expression for the propagator 
\begin{eqnarray}
 G_{\nu}(- \cos s ) = \frac {\Gamma({\frac {d-1}{2} + i\nu} )\Gamma({\frac {d-1}{2} - i\nu} )}{2 (2\pi)^{d/2}} (\sin s)^{-\frac {d-2}{2}} \P_{-\frac 12 +i \nu }^{-\frac {d-2}2}(-\cos s)  \label{leg6i}
\end{eqnarray}
where
$\P_\bbeta^\aalpha(z)$
is the so called "Legendre function on the cut" or Ferrers function of the first kind \cite{bateman} (see Appendix \ref{details1}).
  It is important to  keep in mind that  Ferrers  functions $\P_\beta^\alpha(z)$ and Legendre functions $P_\beta^\alpha(z)$   are holomorphic in
different cut-planes;  as regards Ferrers function this is 
\beq
\Delta_2 = \bC\setminus \{(-\infty,-1]\cup [1,\infty)]\}.
\label{d.20}\endq

\subsection*{Banana integrals}

We are now ready to write the  $n$-loop banana integrals on the sphere with $n+1$ propagators: 
\begin{eqnarray}
I_n(\nu_1,\ldots,\nu_{n+1},d)= \int G^{d}_{\nu_1}(x_0\cdot x)  G^{d}_{\nu_2}(x_0\cdot x) \ldots  G^{d}_{\nu_{n+1}}(x_0\cdot x)\sqrt{g} \, dx;  \label{diagram}\end{eqnarray}
here $x$ varies on  the de Sitter sphere, $x_0$ is a fixed reference point over there and $\sqrt{g} \, dx$ is the rotation invariant measure.
By integrating over the angles we get 
\begin{eqnarray}
&& I_n(\nu_1,\ldots,\nu_{n+1},d)= \cr 
& &  = \frac {2\pi ^{\frac {d}{2}}}{\Gamma \left({\frac {d}{2}}\right)} \int _0^\pi G^{d}_{\nu_1}(-\cos s )  G^{d}_{\nu_2}(-\cos s )  \ldots G^{d}_{\nu_{n+1}}(-\cos s ) (\sin s)^{d-1} ds.\label{bub}
\end{eqnarray}
The aim of this paper is to actually compute the two and  three-lines banana integrals. Here we add some information on the three lines case: this is explicitly written as the integral of the product of three Ferrers function on the interval $(-1,1)$ which is the projection on the plane $\Theta$ of the Euclidean sphere:
\begin{align}
& I_3(\na,\nb,\nc,d)  
&  
= K_3(\na,\nb,\nc,d)
\int_{-1}^{1} \P_{-{1\over 2} +i \na }^{-{d-2 \over 2}}(u )
\P_{-{1\over 2} +i \nb }^{-{d-2 \over 2}}(u )  \P_{-{1\over 2} +i \nc }^{-{d-2 \over 2}}(u ){(1-u^2)^{-\frac{d-2}4}}du,  \label{wm2}
\end{align}
with
\begin{equation}
K_3(\na,\nb,\nc,d) ={
\prod_{j=1}^3\Gamma({d-1\over 2}
- i\nu_j )\Gamma({d-1\over 2} + i\nu_j) )       
\over  2^{2+\frac {3 d}2} \pi ^d 
\Gamma \left(\frac {d}{2}\right) }
 . \label{uci}
\end{equation}
In a previous paper \cite{us} another  integral of three Legendre functions has been computed:
\begin{eqnarray}
h_d(\lambda,\nu, \n) = \int_1^\infty
{P^{-\frac{d-2}{2}}_{-\frac{1}{2} +
i\lambda}(u)P^{-\frac{d-2}{2}}_{-\frac{1}{2} +
i\nu}(u)} P^{-\frac{d-2}{2}}_{-\frac{1}{2} + i \n} (u)\,
{(u^2-1)^{-\frac{d-2}4}} \ du   \label{integralret}
\cr
=
\frac{2^{\frac d 2}}{(4\pi)^{\frac32}
\Gamma\left(\frac{d-1}{2}\right)} \frac{ \ \prod_{\epsilon,\epsilon',\epsilon''=\pm 1}
\Gamma\left(\frac{d-1}{4}
+\frac{i\epsilon\lambda+i\epsilon'\nu +i\epsilon'' \kappa}{2}\right)}
{ \prod_{\epsilon,\epsilon' \epsilon''=\pm 1}
\Gamma\left(\frac{d-1}{2}+i\epsilon \lambda\right)
\Gamma\left(\frac{d-1}{2}
+i\epsilon' \nu\right) \Gamma\left(\frac{d-1}{2}+i\epsilon'' \n\right)}.
\label{theformula}
 \end{eqnarray}
It was far from obvious and  made possible by a  mix of geometrical ideas 
with  analytical and probabilistic tools. 
The steps involved in the proof-computation also gave rise to many interesting quantities having possibly geometrical interpretations that are not yet fully explored in their mathematical and physical consequences. 

An output of the above result is an explicit  K\"all\'en--Lehmann representation of the product of two Wightman two-point functions 
   \begin{eqnarray}
w^d_\lambda(\zeta)w^d_\nu(\zeta) =\int_{-\infty}^{\infty}  \rho_d({\lambda},\nu,\kappa) w^d_\kappa (\zeta) \kappa d\kappa,
\label{KL}\end{eqnarray}
where \begin{eqnarray}
 \rho_d({\lambda},\nu,\kappa)
= \frac{1}{2^{d}{\pi^{{d-1\over 2}}}\kappa
\Gamma\left(\frac{d-1}{2}\right)} \frac{ \ \prod_{\epsilon,\epsilon',\epsilon''=\pm 1}
\Gamma\left(\frac{d-1}{4}
+\frac{i\epsilon\lambda+i\epsilon'\nu +i\epsilon'' \kappa}{2}\right)}
{ \prod_{ \epsilon=\pm 1}
\Gamma
   \left(\frac{i \epsilon \kappa }{2}\right)  \Gamma \left(\frac{1}{2}+\frac{i \epsilon\kappa }{2}\right) \Gamma \left(\frac{d-1}{4}+\frac{i \epsilon \kappa }{2}\right)
   \Gamma \left(\frac{d+1}{4}+\frac{i \epsilon \kappa }{2}\right)} . 
    \cr && 
\label{mf}\end{eqnarray}
Contrary to what happens in flat space, the K\"all\'en--Lehmann 
 weight is here a meromorphic  function of the three mass variables; this fact in particular implies that particle decays that in flat space are forbidden by mass subadditivity,  can take place in the de Sitter universe \cite{us}.

It turns out that the direct evaluation of (\ref{wm2}) is extremely difficult; the geometry of the complex de Sitter universe is less helpful here than it was in the evaluation of the previous integral (\ref{integralret}). The knowledge of the  K\"all\'en--Lehmann representation (\ref{KL}), which is deeply rooted in the harmonic analysis that we have recalled above,  offers to us one opportunity to solve the problem; we are going to describe that construction later, in section \ref{2loop}.


\section{1-loop: the bubble}
\label{1loop}
Let us start with  the already nontrivial two-line case, i.e. the bubble:
\begin{eqnarray}
I(\lambda,\nu,d)  &= & \frac {2\pi ^{\frac {d}{2}}}{\Gamma \left({\frac {d}{2}}\right)} \int _0^\pi G_{\lambda}(-\cos s )  G_{\nu}(-\cos s )(\sin s)^{d-1} ds \label{bub2}\\
&= & 
K(\lambda,\nu,d)
\int_{-1}^{1} \P_{-{1\over 2} +i \lambda }^{-{d-2 \over 2}}(u )
\P_{-{1\over 2} +i \nu }^{-{d-2 \over 2}}(u ) du,  \label{ui0}\\
K_d(\lambda,\nu) & = &
\Gamma({{d-1\over 2}
- i\lambda} )\Gamma({{d-1\over 2} + i\lambda} )\Gamma({{d-1\over 2} - i\nu)\Gamma({{d-1\over 2} + i\nu} )}\over 2 \left(2\sqrt{\pi}\right)^{d}  
\Gamma \left({\frac {d}{2}}\right) 
 . \label{uci}
\end{eqnarray}
To  the best of our knowledge, as simple as it it may look, the integral at the rhs of Eq. (\ref{ui0}) is not listed anywhere in the literature. We will compute it in three different  ways. 
\subsection*{Using the Wronskian}

The method that gives the cleanest result is based on
the properties of
the Wronskian of two Legendre functions. We summarize this result in the following 
\begin{formula} \label{lemma}
\begin{equation}
I(\lambda,\nu,d)  
 =\frac{ \Gamma \left(1-\frac{d}{2}\right) }{2^{d} \pi ^{\frac{d}{2}}(\lambda^2 -\nu^2 ) }\
\left(\frac{ \Gamma \left(\frac{d-1}{2}- i \nu\right) \Gamma \left(\frac{d-1}{2} + i \nu \right)}  {\Gamma \left(\frac{1}{2}- i \nu\right) \Gamma \left(\frac{1}{2} + i \nu \right)}  -\frac{ \Gamma \left(\frac{d-1}{2}- i \lambda\right) \Gamma \left(\frac{d-1}{2} + i \lambda \right)}  {\Gamma \left(\frac{1}{2}- i \lambda\right) \Gamma \left(\frac{1}{2} + i \lambda \right)}
 \right).\ 
\label{finalres}
\end{equation}
\end{formula}
To give a proof of the above formula let us consider two solutions  $u_\nu^\mu$ and $v_\sigma^\mu$  of
the Legendre equation, with $u_\nu^\mu(z)$ standing for either
$\P_\nu^\mu(z)$ or $\Q_\nu^\mu(z)$ and, independently,
$v_\sigma^\mu(z)$ standing for either
$\P_\sigma^\mu(z)$ or $\Q_\sigma^\mu(z)$. Then
\begin{align}
&\int_a^b u_\nu^\mu(z)v_\sigma^\mu(z)(\nu-\sigma)(\sigma+\nu+1)dz =\Big [ (1-z^2)^\half(\sigma+\mu)(\sigma-\mu+1)u_\nu^\mu(z)v_\sigma^{\mu-1}(z)\cr
&-(1-z^2)^\half(\nu+\mu)(\nu-\mu+1)u_\nu^{\mu-1}(z)v_\sigma^\mu(z) \Big ]_a^b\ .
  \label{a.23}\end{align}
This is the main Wronskian equation.
Its derivation  and more details are given
in Appendix \ref{details1} (see also \cite[3.12 (1) p. 169]{bateman}).
If $a= -1$ and $b = 1$, the integral on the l.h.s. converges when
$-1<\Re \mu < 1$. 

In this section we will examine the case when $u_\nu^\mu = \P_\nu^\mu$,
$v_\sigma^\mu = \P_\sigma^\mu$ which is relevant to the subject of this
paper. The cases when $u_\nu^\mu = \P_\nu^\mu$,
$v_\sigma^\mu = \Q_\sigma^\mu$, and when $u_\nu^\mu = \Q_\nu^\mu$,
$v_\sigma^\mu = \Q_\sigma^\mu$ are discussed in Appendix \ref{details1}.

 Let us thus substitute $u_\nu^\mu = \P_\nu^\mu$ and $v_\sigma^\mu = \P_\sigma^\mu$ in (\ref{a.23}):
\begin{align}
&(\nu-\sigma)(\nu+\sigma+1)\int_a^b \P_\nu^\mu(z)\P_\sigma^\mu(z)\,dz =\Big [ (\sigma+\mu)(\sigma-\mu+1)(1-z^2)^\half
\P_\nu^\mu(z)\P_\sigma^{\mu-1}(z)\cr &
-(\nu+\mu)(\nu-\mu+1)(1-z^2)^\half\P_\nu^{\mu-1}(z)\P_\sigma^\mu(z)
\Big ]_a^b 
\label{r.10}\end{align}
and evaluate the r.h.s. of this equation under the
conditions $a= -1$, $b = 1$ and $0< \mu< 1$. To do this it suffices
to evaluate the first term in the r.h.s. of (\ref{r.10}), since the second
term is obtained
by exchanging $\nu$ and $\sigma$ and by a global change of sign.

Since  $0<\mu <1$, as $z \in (0,\ 1)$, $z \rightarrow 1$, 
Eq.  (\ref{d.60}) shows that
\beq
(\sigma+\mu)(\sigma-\mu+1)(1-z^2)^\half
\P_\nu^\mu(z)\P_\sigma^{\mu-1}(z) \sim C (1-z)^{\half - {\mu\over 2}
- {\mu\over 2} +\half} = C(1-z)^{1-\mu} \rightarrow 0\ .
\label{r.55}\endq
As $z \in (-1,\ 1)$, $z \rightarrow -1$  we use Eqs.  (\ref{d.75}) and
(\ref{d.80}) to get 
\begin{align}
& (\sigma+\mu)(\sigma-\mu+1)(1-z^2)^\half
\P_\nu^\mu(z)\P_\sigma^{\mu-1}(z) 
\rightarrow {2\pi^{-1} \sin(\pi\nu)\Gamma(\mu)\Gamma(1-\mu)\over
\Gamma(1+\sigma-\mu)\Gamma(-\sigma-\mu)}\ .
\label{r.80}\end{align}
Putting together the contributions of the two terms in the rhs of
(\ref{r.10}) gives
\begin{equation}
\int_{-1}^1 \P_\nu^\mu(z) \P_\sigma^\mu(z)\, dz =
2\pi^{-1}\Gamma(\mu)\Gamma(1-\mu)\funame_{\P}(\nu,\sigma, \mu),
\label{r.85}\end{equation}
where
\begin{equation}\funame_{\P}(\nu, \sigma, \mu) =
{ {\sin(\pi\sigma)\over\Gamma(1+\nu-\mu)\Gamma(-\nu-\mu)}-
{\sin(\pi\nu)\over\Gamma(1+\sigma-\mu)\Gamma(-\sigma-\mu)} \over (\nu+\sigma+1)(\nu-\sigma)}
.
\label{r.90}\end{equation}
 $\funame_{\P}(\nu, \sigma, \mu)$ 
extends to an entire function of all its arguments,
it is symmetric in $\nu$ and $\sigma$, invariant
under the involution $\nu \rightarrow -\nu-1$ and vanishes at $\mu=0$.
Equation (\ref{r.85}) remains valid, by analytic continuation,
for $-1< \Re \mu <1$.
Combining (\ref{r.85}-\ref{r.90}) with (\ref{uci}) we
obtain the beautiful formula \ref{finalres} for the bubble.

\vskip 10 pt 
Here are a  few  consequences of the above result:
\vskip 20 pt 
\begin{enumerate}
\item The bubble is regular at $d=2$:
\begin{equation}
I(\lambda,\nu,2) =
\frac{\left(\psi
   ^{}\left(\frac{1}{2} - i
   \lambda \right)+\psi
   ^{}\left(\frac{1}{2} + i
   \lambda\right)-\psi
   ^{}\left(\frac{1}{2} - i
   \nu \right)-\psi
   ^{}\left(\frac{1}{2} + i \nu\right)\right)\ }{ 4 \pi 
   \left(\lambda ^2-\nu ^2\right)} \label{1l2m} .
\end{equation}
\item  In odd spacetime dimension the formula becomes
very simple; for instance at $d=3$
\begin{equation}
I(\lambda,\nu,3) =\frac{\lambda  \coth (\pi  \lambda
   )-\nu  \coth (\pi  \nu )}{4 \pi 
   \lambda ^2-4 \pi  \nu ^2}.
  \end{equation}
 \item  At $d=4$ we encounter the first divergence. The Laurent expansion of the formula near $d=4$ gives  
 \begin{eqnarray}
&& I(\lambda,\nu,4) =-\frac{1}{8 \pi ^2 (d-4)}+ \frac{1-\gamma +\log (4 \pi )}{16 \pi
   ^2} \cr && -\frac{\left(4 \lambda ^2+1\right) \left(\psi
  \left(\frac{3}{2}-i \lambda \right)+\psi
  \left(
   \frac{3}{2}+i \lambda\right)\right)-\left(4 \nu
   ^2+1\right) \left(\psi
   \left(\frac{3}{2}-i \nu \right)+\psi
  \left(
   \frac{3}{2}+i \nu\right)\right)}{64 \pi ^2
   \left(\lambda ^2-\nu ^2\right)}
\cr&&+
\ {\rm O}(d-4). \label{Ilambdanu4}
 \end{eqnarray}
\end{enumerate}

\section{The $O(N)$ model: a summary}\label{sec2}
Let $E\cong \mathbb R^N$ be the $N$ dimensional Euclidean vector space with the standard  scalar product denoted by $\langle \cdot\mid \cdot  \rangle$ and $O(N)$ the corresponding orthogonal group. Let us consider a scalar multi-component field 
\begin{align}
 \phi: S_d \longrightarrow E,
\end{align}
where $S_d$ is the (Euclidean) de Sitter sphere of radius $R$.  The radius $R$ and the cosmological constant 
\begin{equation}
    \Lambda=\frac{(d-1)(d-2)}{2R^2}
\end{equation} 
will reappear when necessary (see. Sect.  (\ref{flat})). For the moment we take $R=1$.

The action for $\phi$ is  the quartic $O(N)$-invariant action
\begin{align}
 S[\phi]=\int_{S_d}  \Big[ \Lambda_0+\frac 12 \langle \partial_\mu \phi \mid \partial^\mu \phi \rangle +\frac {m_0^2}2 \langle \phi \mid \phi \rangle +\frac {c_0}4 \langle \phi \mid \phi \rangle^2 \Big] \sqrt {g}\, d^d x,
\end{align}
where $m_0$ and $c_0$ are respectively the bare mass and bare self-coupling constant; $x^\mu$ are local coordinates on the sphere. We included also an extra cosmological  constant $\Lambda_0$ for convenience when renormalizing.

We will compute the effective potential at $d=4$ for the constant configuration 
\begin{align}
 \bar \phi\equiv \varphi\, e_0,
\end{align}
where $\varphi$ is a real constant and $e_0$ is a given vector of norm $1$ in $E$. Of course a nonzero expectation value of $ \phi$  breaks the symmetry down to $O(N-1)$. After choosing  any orthonormal basis $\{e_j\}_{j=0}^{N-1}$ of $E$ whose first element is $e_0$, we may write 
\begin{align}
 \phi=(\varphi+\psi_0)e_0+\sum_{j=1}^{N-1} \psi_j e_j,
\end{align}
so that
\begin{align}
 S[\phi]=& \int_{S_d} \Big[ \Lambda_0+ \frac {m_0^2}2 \varphi^2 +\frac {c_0}4 \varphi^4 \Big]\sqrt {g}\, d^d x +\sum_{a=1}^4 S_a[\psi;\varphi],
\end{align}
where
\begin{align}
 S_1[\psi;\varphi]=& \int_{S_d}  \Big[{m_0^2} \varphi \psi_0 + {c_0} \varphi^3 \psi_0 \Big]\sqrt {g}\, d^d x,\\
 S_2[\psi;\varphi]=& \frac 12 \int_{S_d}  \Big[\sum_{j=0}^{N-1}\partial_\mu \psi_j \partial^\mu \psi_j + (m_0^2+3c_0 \varphi^2) \psi_0^2 + (m_0^2+c_0 \varphi^2) \sum_{j=1}^{N-1}\psi_j^2 \Big] \sqrt {g}\,d^d x, \\
 S_3[\psi;\varphi]=& \int_{S_d}  \Big[c_0 \varphi \psi_0 \sum_{j=0}^{N-1} \psi_j^2 \Big] \, d^d x,\\
 S_4[\psi;\varphi]=& \frac {c_0}4\int_{S_d} \sqrt {g} \Big[\sum_{j=0}^{N-1} \psi_j^2 \Big]^2 \sqrt {g}\, d^d x.
\end{align}
The effective potential ${\cal V}(\varphi)$ is defined as follows
\begin{align}
 \exp (-\Omega_d {\cal V}(\varphi))=\int [\prod_j D\psi_j] \exp (-S[\phi]);
\end{align}
here $\Omega_d$ is the volume of the Euclidean de Sitter spacetime (which is finite) and $[\prod_j D\psi_j]$ is the formal path integral measure.
By construction, $S_1$ does not contribute to the effective potential; at 1-loop, we get 
\begin{align}
 {\cal V}(\varphi)=&  \Lambda_0+ \frac {m_0^2}2 \varphi^2 +\frac {c_0}4 \varphi^4- \frac 1{\Omega_d} \log \int [\prod_j D\psi_j] \exp (-S_2[\phi]),
\end{align}
which we rewrite in the standard form: 
\begin{align}
 {\cal V}(\varphi)=&  \Lambda_0+ \frac {m_0^2}2 \varphi^2 +\frac {c_0}4 \varphi^4 +{\cal V}_0(\varphi)+(N-1) {\cal V}_1(\varphi),
\end{align}
where, for $a=0,1$,
\begin{align}
& {\cal V}_a(\varphi)= - \frac 1{\Omega_d} \log \int [D\Phi] \exp \left(-\frac 12 \int_{M_d} \sqrt {g} \Big[\partial_\mu \Phi \partial^\mu \Phi + M_a^2(\varphi) \Phi^2 \Big] d^d x\right), \\
& M_0^2(\varphi)= m_0^2+3c_0 \varphi^2, \\
&  M_1^2(\varphi)= m_0^2+c_0 \varphi^2.
\end{align}

Several techniques can be used to compute ${\cal V}_a(\varphi)$. 
Instead of looking at it as a function of $\varphi$, it is useful to see it as a function of $M_a^2\equiv z_a$. 
By taking the first derivative w.r.t. $z$ we get the tadpole integral which leads to the Lee-Sciaccaluga equation for determining the effective potential \cite{Lee-Sciaccaluga}.

When we differentiate twice, we get the one mass bubble integral; this is the strategy we want to adopt, since it will allow us to develop methods that will be helpful to compute the two-loop contribution (a research that will be considered elsewhere).
More precisely, if $J(m^2_1,m^2_2)$ is the two-masses bubble, one has
\begin{align}
   \frac {\partial^2}{\partial z^2}{\cal V}(\varphi)=-\frac 12 J(z,z),
\end{align}
from which we can recover the effective potential. As we will review in Sec. \ref{flat}, in the flat case and $d=4-2\epsilon$, we have 
\begin{align}
 J(z,z)=\frac 1{16\pi^2} \left( \frac 1\epsilon - \log \frac {z}{4\pi e^\gamma} \right)+O(\epsilon).\label{jzz}
\end{align}
After integrating twice in $z$ and renormalising, we get the standard result \cite{ColemanW,FordJones}
\begin{align}
 {\cal V}(\varphi)=&  \frac {m^2}2 \varphi^2 +\frac {c_g}4 \varphi^4 +\frac{(m^2+3c_g \varphi^2)^2}{64\pi^2} \log \frac {m^2+3c_g \varphi^2}{\mu^2} \cr
 &+(N-1) \frac {(m^2+c_g \varphi^2)^2}{64\pi^2} \log \frac {m^2+c_g \varphi^2}{\mu^2}, \label{flatpot}
\end{align}
where $m$ and $c_g$ are the renormalised mass and coupling constant, and $\mu^2$ is a scale related to the precise renormalisation conditions. We will now extend this calculation to the de Sitter case.

\section{The effective potential}\label{sec5}
To compute the effective potential at $d=4$, we set $d=4-2\epsilon$. Since $d\nu^2=dm^2$, we have to integrate twice $- I(\nu,\nu,4)/2$ in $d\nu^2=2\nu d\nu$. From \eqref{Ilambdanu4}, we have
\begin{align}
  I(\nu,\nu,4-2\epsilon) =&\frac{1}{16 \pi ^2 \epsilon}+ \frac{1-\gamma +\log (4 \pi )}{16 \pi^2} \cr 
  & -\frac 1{128\pi^2 \nu}\frac {d}{d\nu}\left[\left(4 \nu^2+1\right) \left(\psi \left(\frac{3}{2}-i \nu \right)+\psi \left(\frac{3}{2}+i \nu\right)\right)\right]
\cr&+
\ {\rm O}(\epsilon).
\end{align}
The nontrivial part comes from the second line. A first integration gives:
\begin{align}
 \int \frac 1{8\nu}\frac {d}{d\nu}\left[\left(4 \nu^2+1\right) \left(\psi \left(\frac{3}{2}-i \nu \right)+\psi \left(\frac{3}{2}+i \nu\right)\right)\right]2\nu d\nu=\cr
 =\left(\nu^2+\frac 14\right) \left(\psi \left(-\frac{1}{2}-i \nu \right)+\psi \left(-\frac{1}{2}+i \nu\right)\right)+C,
\end{align}
which, integrated once again becomes
\begin{align}
 C\nu^2+\int \left(\nu^2+\frac 14\right) \left(\psi \left(-\frac{1}{2}-i \nu \right)+\psi \left(-\frac{1}{2}+i \nu\right)\right)d\nu^2. \label{integrale}
\end{align}
An integration by parts then gives
\begin{align}
 \int \left(\nu^2+\frac 14\right)& \left(\psi \left(-\frac{1}{2}-i \nu \right)+\psi \left(-\frac{1}{2}+i \nu\right)\right)d\nu^2=\cr
 &=\frac 12 \left(\nu^2+\frac 14\right)^2 \left(\psi \left(-\frac{1}{2}-i \nu \right)+\psi \left(-\frac{1}{2}+i \nu\right)\right)\cr
 &-\frac i2 \int d\nu \left(\nu^2+\frac 14\right)^2 \left(\psi' \left(-\frac{1}{2}-i \nu \right)-\psi' \left(-\frac{1}{2}+i \nu\right)\right) .
\end{align}
Notice that 
\begin{align}
 \psi'(z)=\zeta(2,z)=\sum_{n=0}^\infty \frac 1{(z+n)^2},\label{hurwitz}
\end{align}
is a Hurwitz zeta function \cite{NIST}. After setting $y:=\nu^2+\frac 14$, we can introduce the function
\begin{align}
 B(y)\equiv&\frac {y^2}4-\frac i2 \int d\nu \left(\nu^2+\frac 14\right)^2 \left(\psi' \left(-\frac{1}{2}-i \nu \right)-\psi' \left(-\frac{1}{2}+i \nu\right)\right) \cr
 =&-\frac 12 \int dy\ y^2 \left(\sum_{n=0}^\infty \frac {2n-1}{(n(n-1)+y)^2} -\frac 1y\right), \label{Bfunction}
\end{align}
where in the second line we have used the series expansion \eqref{hurwitz}.
By using the standard Abel-Plana's formula \cite{AbelPlana} we can write
\begin{align}
\sum_{n=0}^\infty \frac {2n-1}{(n(n-1)+y)^2}=& - \frac 1{2y^2}-4 \int_0^\infty \frac t{e^{2\pi t}-1}\frac {(t^2-y)^2-y}{[(t^2-y)^2+t^2]^2},
\end{align}
so that
\begin{align}
    B(y)=& \frac y3+ \int_0^\infty \frac {2t^4+t^2-1+y(4-2t^2)}{(t^2-y)^2+t^2} \frac {t^3 dt}{e^{2\pi t}-1} +6\int_0^\infty \arctan \left(t-\frac yt\right) \frac {t^2 dt}{e^{2\pi t}-1}\cr
    &+2\int_0^\infty \left(t^2-\frac 12\right) \log [t^2+(t^2-y)^2] \frac {t dt}{e^{2\pi t}-1}, \label{Butile}
\end{align}
where we have conveniently set to zero an integration constant.
With these notations, the (renormalized) effective potential is 
\begin{align}
    {\cal V}_R=&\frac {m_R^2}2 \varphi_R^2 +\frac {c_g}4 \varphi_R^4 +\frac 1{64\pi^2} (m_R^2+3c_g \varphi_R^2-2)^2 \left(\psi \left(-\frac{1}{2}-i \nu_0 \right)+\psi \left(-\frac{1}{2}+i \nu_0\right)\right)\cr
    &+\frac 1{32\pi^2} B({m_R^2+3c_g \varphi_R^2-2})+\frac {N-1}{32\pi^2} B({m_R^2+c_g \varphi_R^2-2}) \cr
    &+\frac {N-1}{64\pi^2} (m_R^2+c_g \varphi_R^2-2)^2 \left(\psi \left(-\frac{1}{2}-i \nu_1 \right)+\psi \left(-\frac{1}{2}+i \nu_1\right)\right)\cr
    & -\frac 1{64\pi^2} \left[(m_R^2+3c_g \varphi_R^2-2)^2+ (N-1)(m_R^2+c_g \varphi_R^2-2)^2 \right] \log {\mu_R^2}. \label{formulazza}
\end{align}
The last line is just the introduction of a reference energy scale, related to the renormalization, while 
\begin{align}
    \nu_0=& \sqrt{m_R^2+3c_g \varphi_R^2-\frac 94}, \\
    \nu_1=& \sqrt{m_R^2+c_g \varphi_R^2-\frac 94}.
\end{align}
$R$ refers to the de Sitter radius and appears through the formula $m_R=mR$, $\mu_R=\mu R$, and $\varphi_R=R\varphi$, $m_R$, $\mu_R$ and $\varphi_R$ being adimensional quantities.\\
We can find an alternative expression for the function $B$ by directly integrating \eqref{Bfunction}:
\begin{align}
  B(y)= -\sum_{n=1}^\infty \Big[ & (2n+1)y+\frac {n(n+1)(2n+1)y}{n(n+1)+y}\cr
  &-2(2n+1)(n+1)n \log \Big( 1+\frac {y}{n(n+1)}\Big)\Big]+\frac {y^2}4+D,
\end{align}
where $D=B(0)$ is a constant that can be determined from \eqref{Butile} as
\begin{align}
    D=-\frac 1{72}+\frac {\gamma-\log(2\pi)}{60}-\frac {3\zeta(3)}{4\pi^2}+4\log(A_3),
\end{align}
where $A_3$ is the third generalized Glaisher-Kinkelin constant (see Appendix \ref{D}).\\
This formula allows us to write the effective potential at any finite value of $\nu$, but it is not suitable for taking the flat limit, for which it is convenient to use \eqref{Butile}.

\section{Flat limit} \label{flat}
Before discussing the flat limit let us recall that   in flat space space the Schwinger function is proportional to a MacDonald function:
\begin{eqnarray}
 G_m(x) =  \frac{1}{(2\pi)^{d }} \int \frac{e^{-ipx}}{{p^2+m^2}}  dp   =
 \frac 1 {(2\pi)^{\frac d 2 }}   \left(\frac{r }{ m}\right)^{1-\frac{d}{2}}  K_{\frac{d}{2}-1}\left( m r \right), \ \ \ r=\sqrt{x^2}.
   \end{eqnarray}
An easy computation in $x$-space \cite{Cacciatori2023a} gives the textbook answer for the bubble:
\begin{eqnarray}
 \int G_{m_1}(x)  G_{m_2}(x)    dx = 
\frac { \Gamma \left(1-{\frac {d}{2}}\right)} {(4\pi)^{{\frac d 2}  }}  
 \frac{  m_2^{{d}-2}-m_1^{{d}-2} }{ m_1^2-m_2^2  } .
  \label{prima deduzione}
\end{eqnarray}
which, when $d=2$, reduces to
\begin{equation}
 I(m_1,m_2,2)=  \frac{\log(m_1)-\log(m_2)}{2 \pi  (m_1^2-m_2^2) } .
\end{equation}
To study  the flat limit of Eq. (\ref{finalres}), we have to restore a generic de Sitter radius $R$ and rescale the masses accordingly. When $m >0$ and $R\rightarrow +\infty$, by using the Stirling formula
\beq
\left |\frac {\Gamma\left ( \frac {d-1}{2} +iRm \right )}{\Gamma\left ( {\frac 12} +iRm \right )} \right |^2 \sim (Rm)^{d-2}\ .
\endq
Hence,
\beq
I(Rm_1,Rm_2,d) \sim  - R^{d-4} \frac {\Gamma\left ( 1-{\frac d2} \right)
  (m_1^{d-2} - m_2^{d-2})}{2^d\pi^{\frac d2}(m_2^2 - m_1^2)}\ \ \ 
(R\rightarrow +\infty)\ .
\endq
This fits exactly  with (\ref{prima deduzione}). 

Once more, we would like to draw the attention of the reader on the non triviality of this result as it expresses the flat limit of an integrated quantity as opposed to the asymptotic properties of the correlation functions which are merely local properties.

Even more interestingly, the same nontrivial result holds true for the effective potential. To show it, we need to compute the behaviour of $B(\nu)$ for $\nu\to\infty$. This is obtained from \eqref{Butile} by noticing that the integral in the first line goes to zero and using that
\begin{align}
    \int_0^\infty \frac {t^k dt}{e^{2\pi t}-1}=\frac {k!}{(2\pi)^{k+1}}\zeta(k+1).
\end{align}
Therefore,
\begin{align}
    B\left(\nu^2+\frac 14\right)=\frac 13 \nu^2 -\frac 1{15} \log \left( \nu^2+\frac 14 \right)+\frac 1{12}- \frac {3\zeta(3)}{4\pi^3}+\ldots,
\end{align}
where the dots stay for terms vanishing for $\nu\to\infty$.\\
On the other hand, for $|\arg(x)|\leq \pi-\delta$, $\delta>0$, one has
\begin{align}
    \psi(x)=\log x-\frac 1{2x} -\sum_{j=1}^m \frac {B_{2j}}{2j} \frac 1{x^{2j}}+O\left( \frac 1{x^{2m+2}} \right),
\end{align}
where $B_j$ are the Bernoulli numbers. Since $B_2=\frac 16$ and $B_4=-\frac 1{30}$, it follows that 
\begin{align}
    \frac 12 \left(\nu^2+\frac 14\right)^2 \left(\psi \left(-\frac{1}{2}-i \nu \right)+\psi \left(-\frac{1}{2}+i \nu\right)\right)=&
    \frac 12 \left( \nu^2+\frac 14 \right)^2 \log \left( \nu^2+\frac 14 \right)\cr &+\frac 13\nu^2+\frac 1{20}+\ldots.
\end{align}
In this limit we have
\begin{align}
    {\cal V}_R=&\frac {m_R^2}2 \varphi_R^2 +\frac {c_g}4 \varphi_R^4 +\frac 1{32\pi^2} \Bigg[\frac {(m_R^2+3c_g \varphi_R^2-2)^2}2 \log \frac {m_R^2+3c_g \varphi_R^2-2}{\mu_R^2} \cr
    &+\frac 23 (m_R^2+3c_g \varphi_R^2-2)-\frac 1{15} \log (m_R^2+3c_g \varphi_R^2-2)-\frac 1{30}-\frac {3\zeta(3)}{4\pi^3}+\ldots\Bigg]\cr
    &+\frac {N-1}{32\pi^2} \Bigg[\frac {(m_R^2+c_g \varphi_R^2-2)^2}2 \log \frac{m_R^2+c_g \varphi_R^2-2}{\mu_R^2} \cr
    &+\frac 23 (m_R^2+c_g \varphi_R^2-2)-\frac 1{15} \log (m_R^2+c_g \varphi_R^2-2)-\frac 1{30}-\frac {3\zeta(3)}{4\pi^3}+\ldots\Bigg].
\end{align}
The potential in the flat limit is defined by ${\cal V}=\lim_{R\to\infty} R^{-4}{\cal V}_R$ and gives exactly \eqref{flatpot}. For very large but finite $R$, we can write
\begin{align}
    {\cal V}_\Lambda=&\frac {m^2}2 \varphi^2 +\frac {c_g}4 \varphi^4 +\frac 1{64\pi^2} \Bigg[(m^2+3c_g \varphi^2)^2 \log \frac {m^2+3c_g \varphi^2}{\mu^2} \cr
    & +(N-1) (m^2+c_g \varphi^2)^2 \log \frac {m^2+c_g \varphi^2}{\mu^2} \Bigg] \cr
    &-\frac {\Lambda}{48\pi^2} \Bigg[ (m^2+3c_g \varphi^2)\Bigg( \log \frac {m^2+3c_g \varphi^2}{\mu^2}+\frac 1{6}\Bigg) \cr 
    &\phantom {-\frac {\Lambda}{48\pi^2} \Bigg[}\ + (N-1)(m^2+c_g \varphi^2)\Bigg( \log \frac {m^2+c_g \varphi^2}{\mu^2}+\frac 1{6}\Bigg) \Bigg]\cr
    &+\frac {\Lambda^2}{144\pi^2}\Bigg[
    -\frac 1{30} \log \frac{3(m^2+3c_g \varphi^2)}{\Lambda}+\log \frac {m^2+3c_g \varphi^2}{\mu^2}+\frac {49}{60}-\frac {3\zeta(3)}{4\pi^3} \cr
    &+(N-1)\Bigg(
    -\frac 1{30} \log \frac{3(m^2+c_g \varphi^2)}{\Lambda}+\log \frac {m^2+c_g \varphi^2}{\mu^2}+\frac {49}{60}-\frac {3\zeta(3)}{4\pi^3}\Bigg)\Bigg]\cr
    &+O(\Lambda^3),
\end{align}
where  $\Lambda=3R^{-2}$ is the cosmological constant. The first two lines are the standard Coleman-Weinberg effective potential \cite{ColemanW,FordJones}, while the remaining terms provide the perturbative corrections in the cosmological constant, up to order two. The exact expression is given by \eqref{formulazza}, which in terms of the full dimensional quantities is 
\begin{align}
    {\cal V}_\Lambda=&\frac {m^2}2 \varphi^2 +\frac {c_g}4 \varphi^4 +\frac 1{64\pi^2} (m^2+3c_g \varphi^2-\frac 23\Lambda)^2 \left(\psi \left(-\frac{1}{2}-i \nu_0 \right)+\psi \left(-\frac{1}{2}+i \nu_0\right)\right)\cr
    &+\frac {\Lambda^2}{288\pi^2} B\left(3 \frac {m^2}{\Lambda}+9c_g \frac {\varphi^2}{\Lambda}-2\right)+\frac {(N-1)\Lambda^2}{288\pi^2} B\left(3 \frac {m^2}{\Lambda}+3c_g \frac {\varphi^2}{\Lambda}-2\right) \cr
    &+\frac {N-1}{64\pi^2} (m^2+c_g \varphi^2-\frac 23\Lambda)^2 \left(\psi \left(-\frac{1}{2}-i \nu_1 \right)+\psi \left(-\frac{1}{2}+i \nu_1\right)\right)\cr
    & -\frac 1{64\pi^2} \left[(m^2+3c_g \varphi^2-\frac 23\Lambda)^2+ (N-1)(m^2+c_g \varphi^2-\frac 23\Lambda)^2 \right] \log \frac {3\mu^2}{\Lambda}, \label{esatto}
\end{align}
with
\begin{align}
    \nu_0=& 3\sqrt{ \frac {m^2}{3\Lambda}+c_g \frac {\varphi^2}{\Lambda}-\frac 14}, \\
    \nu_1=& 3\sqrt{\frac {m^2}{3\Lambda}+c_g \frac {\varphi^2}{3\Lambda}-\frac 14}.
\end{align}

It is interesting to compare this result with the one for anti de Sitter, \cite{HSUAdS}. In that case, the perturbative expansion in the cosmological constant $\Lambda$ of the effective potential contains also terms of order $\sqrt {|\Lambda|}$. This is strictly related to the invariance of the Legendre functions of the first kind $P^\mu_{-\frac 12-\nu}$ under the $\nu\to-\nu$, which is not true for the functions of the second kind.

\section{2-loop: the watermelon}

\label{2loop}

 For the sake of comparison let us recall at first the flat space formula for the watermelon:
\begin{align}
I_3(m_1,m_2,m_3,d)& =  -2^{1-2 d} \pi ^{1-d} \Gamma (2-d) (-S(\a,\b,\c))^{\frac{d-3}{2}}\cr 
&+ \frac{ (\a\b)^{d-4}
   \left(\a^2+\b^2-\c^2\right) \,
   _2F_1\left(1,2-\frac{d}{2};\frac{3}{2};M_{123}^2\right)}{4^{d} \pi ^{d-2}(\cos (\pi  d)-1)
   \Gamma \left(\frac{d}{2}-1\right) \Gamma
   \left(\frac{d}{2}\right)}
   \cr 
&+  \frac{ (\b\c)^{d-4}
   \left(-\a^2+\b^2+\c^2\right) \,
   _2F_1\left(1,2-\frac{d}{2};\frac{3}{2};M_{231}^2\right)}{4^{d} \pi ^{d-2}(\cos (\pi  d)-1)
   \Gamma \left(\frac{d}{2}-1\right) \Gamma
   \left(\frac{d}{2}\right)}  
   \cr 
&+  \frac{ (\a\c)^{d-4}
   \left(\a^2-\b^2+\c^2\right) \,
   _2F_1\left(1,2-\frac{d}{2};\frac{3}{2};{M_{312}^2}\right)}{4^{d} \pi ^{d-2}(\cos (\pi  d)-1)
   \Gamma \left(\frac{d}{2}-1\right) \Gamma
   \left(\frac{d}{2}\right)}.  \cr & \label{bf2}
\end{align}
where 
\begin{equation} M_{ijk} = \left(\frac{m_i^2+m_j^2-m_k^2}{2 m_i m_j }\right)
\end{equation}
and 
\begin{equation}
S(\a,\b,\c)=\a^4+\b^4+\c^4-2 \a^2 \b^2-2 \a^2 \c^2- 2 \b^2 \c^2 \label{311}
   \end{equation}
is the Symanzik polynomial. The above formula is valid when no  mass is bigger than the sum of the other two; this happens if and only if  the Symanzik polynomial is negative. It may be obtained by solving an appropriate differential equation for the diagram \cite{FordJones} or else by a direct calculation in position space which is indeed the easier and shorter way to get it \cite{Cacciatori2023a}. A similar formula holds for $S>0$ \cite{Cacciatori2023a}.

As regards de Sitter sphere the situation is trickier. At the moment there is no substitute for the method of differential equations which seems to be well-adapted only to flat space but useless otherwise. We are left with $x$-space calculations.

Faced directly, the integral at the rhs of (\ref{wm2}) is significantly more challenging than the already difficult (\ref{theformula}) and at the moment the only way we found to evaluate it  makes indeed use of Eq.  (\ref{theformula}): by singling out one of the masses, say $\nu_3$, we get an integral representation of the two-loop watermelon as a superposition of one-loop diagrams:
\begin{eqnarray}
&& I_3(\nu_1,\nu_2,\nu_3,d)= \int G_{\nu_1}(x_0\cdot x)  G_{\nu_2}(x_0\cdot x)  G_{\nu_3}(x_0\cdot x) \sqrt g dx  =\label{opopo}\cr&& =
 \int \rho (\nu_1,\nu_2,\kappa ) I_2(\kappa,\nu_3,d) \kappa d\kappa  =   
 {I}_3^{(1)} (\nu_1,\nu_2,\nu_3,d)- I_3^{(2)}(\nu_1,\nu_2,\nu_3,d).\label{general}
 \end{eqnarray}
where
\begin{eqnarray}
&& {I}_3^{(1)} (\nu_1,\nu_2,\nu_3,d)=
\frac{ \left(\frac{ \Gamma \left(\frac{d-1}{2}- i \nu_3\right) \Gamma \left(\frac{d-1}{2} + i \nu_3 \right)}  {\Gamma \left(\frac{1}{2}- i \nu_3\right) \Gamma \left(\frac{1}{2} + i \nu_3 \right)}   
 \right)}{ 2^{d} \pi^{\frac{d-1}{2}}\Gamma
   \left(\frac{d-1}{2}\right)}
  \frac{\Gamma \left(1-\frac{d}{2}\right)}{(4 \pi) ^{\frac d 2}}
 {\cal A}(\na,\nb,\nc,d)  \label{p09o} \\ &&  {\cal A}(\na,\nb,\nc,d)=\int 
   \frac{ \prod_{\epsilon,\epsilon',\epsilon''}
\Gamma\left(\frac{d-1}{4}
+\frac{i\epsilon'\nu_1+i\epsilon'\nu_2+i\epsilon'' \kappa}{2}\right)}
{\prod_{\epsilon}
\Gamma
   \left(\frac{i \epsilon \kappa }{2}\right)  \Gamma \left(\frac{1}{2}+\frac{i \epsilon\kappa }{2}\right) \Gamma \left(\frac{d-1}{4}+\frac{i \epsilon \kappa }{2}\right)
   \Gamma \left(\frac{d+1}{4}+\frac{i \epsilon \kappa }{2}\right)} \frac {d\kappa} { (\kappa^2 -\nu_3^2 ) }\
\cr &&  \label{calA}
\end{eqnarray}
and 
\begin{align}
&{I}_3^{(2)} (\nu_1,\nu_2,\nu_3,d)=
\frac{ 1}{ 2^{d} \pi^{\frac{d-1}{2}}\Gamma
   \left(\frac{d-1}{2}\right)}
  \frac{\Gamma \left(1-\frac{d}{2}\right)}{(4 \pi) ^{\frac d 2}}{\cal B}(\na,\nb,\nc) \label{p09oi}
 \\ &  {\cal B}(\na,\nb,\nc)= \int  
   \frac{ \frac {{ \Gamma \left(\frac{d-1}{2}- i \kappa\right) \Gamma \left(\frac{d-1}{2} +i \kappa \right)}   } { {\Gamma \left(\frac{1}{2}- i  \kappa\right) \Gamma \left(\frac{1}{2} +i \kappa\right)} }\prod_{\epsilon,\epsilon',\epsilon'' }
\Gamma\left(\frac{d-1}{4}
+\frac{i\epsilon'\nu_1+i\epsilon'\nu_2+i\epsilon'' \kappa}{2}\right)}
{\prod_{\epsilon }
\Gamma
   \left(\frac{i \epsilon \kappa }{2}\right)  \Gamma \left(\frac{1}{2}+\frac{i \epsilon\kappa }{2}\right) \Gamma \left(\frac{d-1}{4}+\frac{i \epsilon \kappa }{2}\right)
   \Gamma \left(\frac{d+1}{4}+\frac{i \epsilon \kappa }{2}\right)}
   \frac {   d\kappa} { (\kappa^2 -\nu_3^2 ) }.\cr&
\label{calB }\end{align}
${I}_3^{(1)} (\nu_1,\nu_2,\nu_3,d)$  and ${I}_3^{(2)} (\nu_1,\nu_2,\nu_3,d)$ are  symmetric only in the exchange of $\nu_1$ and $ \nu_2$. To work with symmetric expressions we might replace them by their totally symmetric counterparts but this is not really helpful. 

\subsection*{First term}
Let us consider the first term ${I}_3^{(1)} (\nu_1,\nu_2,\nu_3,d)$; the change of variables\footnote{In the following we will use either $(\na,\nb,\nc)$ or $(x,y,w)$ and either $d$ or $\delta(d)$ interchangeably. } \begin{equation}
  s = \frac{i \kappa} 2 ,\ \ \   \mmu = \frac {d-1}4, \ \ \  x  =  \frac {i\nu_1} 2,\ \  \ y =    \frac {i\nu_2} 2,\ \ \ \ w =    \frac {i\nu_3} 2, \ \ \ \ u=x+y \label{pol}
\end{equation}allows to rewrite the integral as follows:
\begin{align}
& {\cal A}(x,y,w,d) 
= 
   \int_{-i\infty}^{i\infty}    \frac{ds }{2 i }\
\prod_{\epsilon, \epsilon' ,\epsilon''=\pm}   \frac{  { \ 
\Gamma\left(\epsilon s +\mmu
+\epsilon' x+i\epsilon''y\right) \Gamma \left( \epsilon s+w\right) }
 }   {   \Gamma
   \left(\epsilon s\right)    \Gamma \left(\frac{1}{2}+\epsilon s\right)  \Gamma \left(\mmu + \epsilon s \right) \Gamma \left(\mmu + \frac 12 + \epsilon s\right) \Gamma \left( \epsilon s+w+1\right)}  .
    \label{bubble1a} 
\end{align}
 There   are  poles   on the integration path at 
\begin{eqnarray} s =-\mmu + \frac{\pm i \nu_1\pm i \nu_2}{2}  -n, \ \ \ \ \ s= - \frac{ i \nu_3}{2}  -n, \\s= - \frac{ i \nu_3}{2}  +n,\ \ \ \ \ \ s = \mmu + \frac{\pm i \nu\pm i \lambda}{2}  + n.\end{eqnarray}
By integrating along the imaginary axis (with suitable indentations)   
the result may be  expressed as  a linear combination of four hypergeometric functions 
\begin{equation}
{I}_3^{(1)} (x,y,z,d)=\sum_{\epsilon , \epsilon' =\pm 1} A_{d}(\epsilon x,\epsilon' y,w )\label{i1}
 \end{equation} 
where
\begin{align}& A_{d}(x,y,w)= a_{d}(x,y,w)
 \times \cr & 
 \, _9F_8\left(\begin{array}{l}
   { 2 \delta, u+\frac{1}{2},u+1 ,\delta +u+1,2 \delta +2 x,2 \delta +2 y,2
   \delta +2 u,\delta -w+u,\delta +w+u}\cr {
   2 x+1,2 y+1,2 u+1,\delta +u,2 \delta +u,2 \delta+
   u+\frac{1}{2}},\delta -w+u+1,\delta+ w+u+1\end{array};1\right),\cr & \label{ad}
\\& a_{d}(x,y,w)= -
\frac{\scriptstyle{4^{-2 \delta -3} \pi ^{-4 \delta -\frac{1}{2}} \Gamma \left(\frac{1}{2}-2 \delta \right) \cos (2 \pi 
   w)  \Gamma (2 \delta -2 w) \Gamma (2 w+2 \delta ) \Gamma (2 x+2 \delta )
   \Gamma (2 y+2 \delta )}}{\scriptstyle{\sin (2 \pi  x) \sin (2 \pi  y)\Gamma (2 x+1) \Gamma (2 y+1) \left(w^2-(\delta +u)^2\right) \Gamma (-2u-2
   \delta ) \Gamma (2 u+4 \delta )}}.
\end{align}
The hypergeometric series in (\ref{ad}) converges absolutely for $
3 -4\delta = 4-d>0$. 

It is possible to proceed to a simplification of the above expressions by observing that in each term of the hypergeometric series (\ref{ad}) the following product of Pochhammers reduces nicely to a rational function: 

\begin{align}
    \frac{(u+\delta +1)_n (u-w+\delta )_n (u+w+\delta )_n}{(u+\delta )_n (u-w+\delta +1)_n
   (u+w+\delta +1)_n} = \frac{((\delta +u)^2-w^2)  \left(\frac{1}{\delta +n+u+w}+\frac{1}{\delta
   +n+u-w}\right)}{2 (\delta +u)}.
\end{align}
We obtain in this way a formula consisting of eight terms $ {}_7F_6$:
\begin{equation}
{I}_3^{(1)} (x,y,z,d)=\sum_{\epsilon , \epsilon' =\pm 1} A'_{d}(\epsilon x,\epsilon' y,\epsilon''w )
\label{i1bis}
 \end{equation} 
where
\begin{align}& A'_{d}(x,y,w)= a'_{d}(x,y,w)
 \times \cr & 
\, _7F_6\left(\begin{array}{l} 2\delta,u+\frac{1}{2},u+1,\delta +u-w,2 \delta +2 u,2 \delta +2 x,2 \delta +2 y \cr 2
   u+1,2 x+1,2 y+1,\delta +u-w+1,2 \delta +u,2 \delta +u+\frac{1}{2}\end{array} ;1\right)\cr & \label{add}
\\ & a'_{d}(x,y,w)= \cr& 
= \frac{2^{-4 \delta -7} \pi ^{-4 \delta -\frac{1}{2}} \Gamma \left(\frac{1}{2}-2 \delta
   \right) \cos (2 \pi  w) \Gamma (2 \delta -2 w) \Gamma
   (2 w+2 \delta ) \Gamma (2 x+2 \delta ) \Gamma (2 y+2 \delta )}{\sin (2 \pi  x) \sin (2 \pi  y) (\delta +u) \Gamma (2
   x+1) \Gamma (2 y+1) \Gamma (-2 u-2 \delta ) \Gamma (2 u+4 \delta ) (\delta +u-w)}. \cr &
\end{align}
This simplification has not  worsened the convergence: (\ref{add}) converges absolutely under the same conditions: $
 4\delta-3 =- d+4 >0$.

\subsection*{Second term}
As regards the second term, we have 
\begin{eqnarray}
{\cal B}(x,y,w,d)  &=&  \int \frac {ds}{2 i}
   \prod_{\epsilon,\epsilon',\epsilon''=\pm 1}\frac{2^{d-2} \Gamma \left(\epsilon s+w\right)  \Gamma \left(\epsilon s +\delta +\epsilon'x+\epsilon''y\right)}
   {\Gamma \left(\epsilon s  \right) \Gamma \left(\epsilon s+\frac{1}{4}\right) \Gamma \left(\epsilon s+\frac{1}{2}\right) \Gamma \left(\epsilon s+\frac{3}{4}\right)  \Gamma \left(\epsilon s+w+1\right)}
\cr&
\label{bubble2}
\end{eqnarray}
Again,  the result may be expressed as a  linear combination of four hypergeometric functions:
\begin{eqnarray}
 I_3^{(2)}=\sum_{\epsilon , \epsilon' =\pm 1} B_{d}(\epsilon x,\epsilon' y,w )
 \label{i2} \end{eqnarray}
where
\begin{align}
& B_{d}(x,y,w)=b_{d}(x,y,w) \cr 
 & \times  \, _7F_6\left(\begin{array}{c} 2 \delta ,\delta +u+1,\delta -w+u,\delta +w+u,2 \delta +2 x,2 \delta +2 y,2 \delta +2 u\cr 2
   x+1,2 y+1,2 x+2 y+1,\delta +u,\delta -w+u+1,\delta +w+u+1\end{array} ;1\right) \label{bd}
\\ & \cr & b_{d}(x,y,w)=\frac{2^{-8 \delta -4 u-7} \pi ^{1-4 \delta } \Gamma \left(\frac{1}{2}-2 \delta \right) \csc (2 \pi  x) \csc (2 \pi  y) \csc (2 \pi  u)\Gamma (2
   x+2\delta )   \Gamma (2 y+2\delta )}{\left((\delta +u)^2-w^2\right)\Gamma (2 x+1) \Gamma (2
   y+1) \Gamma (2 u+1)  \Gamma (-4 u -4\delta ) \Gamma \left(2 u+2
   \delta +\frac{1}{2}\right)}. \cr & 
\end{align}
The hypergeometric series in (\ref{bd}) converges absolutely for $ 
4 -8\delta = 2(3-d)>0.$
Proceeding exactly as before we may find a simplification of the above formula:
\begin{equation}
{I}_3^{(2)} (x,y,z,d)=\sum_{\epsilon , \epsilon',\epsilon'' =\pm 1} B'_{d}(\epsilon x,\epsilon' y,\epsilon''w )
\label{ii2}
 \end{equation} 
where
\begin{align}
& B'_{d}(x,y,w)=b'_{d}(x,y,w)   \, \, _5F_4\left(\begin{array}{l} 2 \delta ,\delta +u+w,2 \delta +2 u,2 \delta +2 x,2 \delta +2 y\cr 2 u+1,2 x+1,2 y+1,\delta +u+w+1\end{array};1\right)\label{bdd}
\\ & \cr & b'_{d}(x,y,w)=-\frac{2^{-6-4 u-8 \delta } \pi ^{1-4 \delta } \csc (2 \pi  x) \csc (2 \pi  y) \csc (2 \pi 
   u) \Gamma \left(\frac{1}{2}-2 \delta \right) \Gamma (2 x+2 \delta ) \Gamma (2 y+2 \delta
   )}{(w+u+\delta ) \Gamma (1+2 x) \Gamma (1+2 y) \Gamma (1+2 u) \Gamma (1-4 u-4 \delta
   ) \Gamma \left(\frac{1}{2}+2 u+2 \delta \right)}. \cr & 
\end{align}
Again, the simplification has not  worsened the convergence: (\ref{bdd}) converges absolutely for $
4-8 \delta =2(3-d)>0$.

In conclusion, ${I}_3^{(1)}$ is the most regular of the two terms; it starts diverging at $d=4$. The second term diverges already at $d=3$. This means  that no compensation among the two terms may be expected. The only exception is at $d=2$: in that case the watermelon is finite; the two terms have to compensate each other to render the divergence in the coefficient $\Gamma \left(1-\frac{d}{2}\right)$  harmless.

 \section{Odd dimensions $d\leq 1$}
\label{d1} In odd integer dimension $d\leq 1$ the result simplifies a great deal because the hypergeometric  series reduce to finite sums.
Let us examine the simplest case $d=1$; our general formulae give 
\begin{eqnarray}
I^{(1)}_3(\nu_1,\nu_2,\nu_3,1)  &= &\frac{\coth (\pi  \nu_1 ) \coth (\pi  \nu_3 ) \left(\nu_2 ^2-\nu_1 ^2-\nu_3 ^2\right)}{4 \nu_1  \nu_3  (\nu_1
   -\nu_2 -\nu_3 ) (\nu_1 +\nu_2 -\nu_3 ) (\nu_1 -\nu_2 +\nu_3 ) (\nu_1 +\nu_2 +\nu_3 )}+\cr &+&\frac{\coth (\pi  \nu_2 ) \coth
   (\pi  \nu_3 ) \left(\nu_1 ^2-\nu_2 ^2-\nu_3 ^2\right)}{4 \nu_2  \nu_3  (\nu_1 -\nu_2 -\nu_3 ) (\nu_1 +\nu_2 -\nu_3
   ) (\nu_1 -\nu_2 +\nu_3 ) (\nu_1 +\nu_2 +\nu_3 )}\cr  \label{ref0} &&
 \end{eqnarray}
and 
\begin{eqnarray}
I_3^{(2)}(\nu_1,\nu_2,\nu_3,1)  &= & \frac{-\coth (\pi  \nu_1 ) \coth (\pi  \nu_2 ) \left(\nu_3 ^2-\nu_1 ^2-\nu_2 ^2\right)}{4 \nu_1  \nu_2  (\nu_1 -\nu_2
   -\nu_3 ) (\nu_1 +\nu_2 -\nu_3 ) (\nu_1 -\nu_2 +\nu_3 ) (\nu_1 +\nu_2 +\nu_3 )}\cr &-&\frac{1}{2 (\nu_1 -\nu_2 -\nu_3 )
   (\nu_1 +\nu_2 -\nu_3 ) (\nu_1 -\nu_2 +\nu_3 ) (\nu_1 +\nu_2 +\nu_3 )}. \label{i2}\cr&&
 \end{eqnarray}
$I_3^{(2)}(\nu_1,\nu_2,\nu_3,1)$ contains a term which is totally symmetric w.r.t. the mass parameters and a term which is symmetric only  w.r.t.  the exchange of $\nu_1$ and $\nu_2$. 

Subtracting  $I_3^{(2)}$ from $I_3^{(1)}$  reestablishes the global symmetry of the diagram w.r.t. the three mass parameters.  
Since $I_3^{(1)}$  do not include  a totally symmetric term it  can be fully deduced from the knowledge of  $I_3^{(2)}$.

The above situation is generic in  odd negative dimension: the first addendum contains the terms proportional to $\coth \pi \nu_3$ and and the second term contains a totally symmetric contribution $ S(\nu_1,\nu_2,\nu_3,d)$ which is just a product of poles. 

Both the symmetric and the non symmetric contributions admit expansions in partial fractions.  The totally symmetric term has  indeed a very simple expression:
 \begin{eqnarray} 
 S(\nu_1,\nu_2,\nu_3,d) &=&  \frac{ \cos({2\pi i \delta})\pi ^{-4 \mmu } \Gamma (1-4 \mmu ) } {32 } \prod_{\epsilon,\epsilon'=\pm} \frac{\Gamma \left(\mmu + \frac{i \nu_1 }{2} +\frac{i \epsilon \nu_2 }{2}+\frac{i \epsilon'\nu_3 }{2}\right)}
{\Gamma  \left(1-\mmu + \frac{i \nu_1 }{2} +\frac{i \epsilon \nu_2 }{2}+\frac{i \epsilon'\nu_3 }{2}\right)} \label{sii};
 \end{eqnarray}
it amounts to a product of simple poles. The non-symmetric term 
is also  proportional to a product of simple poles but the  polynomial at the numerator becomes more and more cumbersome as the dimension grows (negative).
\vskip 10 pt
A straightforward derivation of  $I_3(\nu_1,\nu_2,\nu_3,1)$   by direct integration of Eq. (\ref{opopo}) points towards a fully symmetric  expression: in spacetime dimension  $d=1$ the Schwinger function 
is simply 
\begin{eqnarray}
 G^1_{\lambda}(- \cos s ) = \frac{{\Gamma (-i \nu ) \Gamma (i
   \nu)} \cosh \left(\lambda (\pi- s)
\right)}{2
   \pi }\label{leg6}
\end{eqnarray}
and therefore
 \begin{eqnarray}
 && I(\nu_1,\nu_2,\nu_3,1) 
= 2 \int _0^\pi G^1_{\nu_1}(-\cos s )  G^1_{\nu_2}(-\cos s )  G^1_{\nu_3}(-\cos s )  ds \cr  
&&= \frac{ \frac{\sinh (\pi  (\na +\nb +\nc
   ))}{\na +\nb +\nc } +\frac{\sinh (\pi  (\na +\nb-\nc ))}{\na +\nb -\nc
   }+\frac{\sinh (\pi  (\na -\nb +\nc ))}{\na -\nb +\nc }+\frac{\sinh (\pi  (\na
   -\nb -\nc ))}{\na -\nb -\nc }}{16 \,\na  \nb  \nc \sinh(\pi  \na ) \sinh(\pi  \nb ) \sinh(\pi  \nc ) }.
 \end{eqnarray} 
 In the special case when all the masses are equal the formula reduces to
  \begin{eqnarray}
 && I(\nu,\nu,\nu,1) =
\frac{5 +\cosh (2 \pi  \nu ) }
{24  \nu ^4\sinh(\pi  \nu )^2}.
 \end{eqnarray}
 Studying the case of three particles with the same mass {\em ab initio} is not easier; it only hides the underlying beautiful structure.
 
Rescaling the masses $\nu\to \nu R$ and taking the limit $R\to\infty$  gives  the correct result in flat space, as obtained by taking $d=1$ in Eq. (\ref{bf2}):
\begin{equation}
I(m_1,m_2,m_3,1) = \frac{1}{4 m_1 m_2 m_3  (m_1+m_2+m_3)};
\end{equation}
we stress again that this  result is non trivial, being the flat limit of an integrated quantity and not just the asymptotic behaviour of a correlation function close to a given point.

Also the above symmetric structure in four terms is generic in odd dimension $d\leq 1$:

 \begin{eqnarray}
 I(\nu_1,\nu_2,\nu_3,d)  
= \sum_{\epsilon,\epsilon'=\pm } \frac{\sinh (\pi  (\na
   +\epsilon \nb +\epsilon' \nc  )) P_d(\na,\epsilon \nb ,\epsilon' \nc) }{\na  \nb  \nc \sinh(\pi  \na ) \sinh(\pi  \nb ) \sinh(\pi  \nc )  Q_d(\na,\epsilon \nb ,\epsilon' \nc)}
 \end{eqnarray} 
where $P_d$ and $Q_d$ are polynomials.
The polynomial at the denominator is always a product of monomials and the result admits a simple partial fraction expansion. We will not elabotrate further on this point.

\section{Spacetime dimension $d=2$}
In dimension $d=2$  the  watermelon is finite. However, as for  the 1-loop diagram  (see Eq. (\ref{1l2m}))  the two terms in Eq. (\ref{general}) are multiplied by the coefficient  $\Gamma \left(1-\frac{d}{2}\right) $ that becomes singular at $d=2$ and a limit procedure is needed. Only in this particular two-dimensional case  ${I}_3^{(1)}$ and ${I}_3^{(2)}$ 
 compensate each other.

To extract the finite result we make use of the elementary identity
\begin{eqnarray}
&& I_3(x,y,z,2)= \left.\frac \partial {\partial d}\left((d-2)I_3(x,y,z,d)\right)\right|_{d=2} = \cr \cr  &&= \left.\partial_d\left((d-2)\sum_{\epsilon , \epsilon',\epsilon''=\pm } (A'_{d}(\epsilon x,\epsilon' y, \epsilon'' w ) -B'_{d}(\epsilon x,\epsilon' y,\epsilon'' w ))\right)\right|_{d=2} .\label{general2}
 \end{eqnarray}
 It is useful to rewrite the typical term in the above expression as follows: 
 \begin{align}&
 A'_d(x,y,w)-B'_d(x,y,w)=  
 a'_{d}(x,y,w)\left (
  \, _7F_6\left(\ldots;1\right) - \frac{b'_{d}(x,y,w)}
{a'_{d}(x,y,w)}
  \, _5F_4\left(\ldots;1\right)\right) .
\end{align}
The term in parentheses at the r.h.s. tends to zero when $d\to 2$. This is precisely the compensation we were expecting. Taking the limit we get 
\begin{align}
& T(x,y,w)=\left.\partial_d\left((d-2)(A'_{d}(x, y,w ) -B'_{d}(x,y,w ))\right)\right|_{d=2} \cr 
& =  c_2(x,y,w ) 
\left.\left( 
\frac \partial {\partial d}
  \,{} _7F_6\left(\ldots;1\right) - 
  \frac \partial {\partial d}\, {} _5F_4\left(\ldots;1\right)
-  {}  _5F_4\left(\ldots;1\right)\ \frac \partial {\partial d} \frac{b'_{d}(x,y,w)}
{a'_{d}(x,y,w)}\right)\right|_{d=2}
 \end{align}
 where $c_{2}(x,y,w) =\lim_{d\to 2} \ (d-2) a'_d(x,y,w).$ All the terms together provide the following finite expression:
 \begin{eqnarray}
&& I_3(x,y,z,2)= \sum_{\epsilon , \epsilon',\epsilon''=\pm } T(\epsilon x,\epsilon' y, \epsilon'' w ) \label{221}
 \end{eqnarray}
where
\begin{align}
   & T(x,y,w) =\frac{\left(\pi  \cot (2 \pi
    (x+y))-\psi\left(\frac{1}{2}-2
   w\right)-\psi \left(\frac{1}{2}+2 w\right)+2
   \psi (1+2 x+2 y)\right)}{32 \sqrt{\pi } \sin
   (2 \pi  x) \sin (2 \pi  y) \Gamma
   \left(\frac{1}{2}-2 x-2 y\right)} \cr & \times
   \frac{ \Gamma \left(\frac{1}{2}+2 x\right) \Gamma
   \left(\frac{1}{2}+2 y\right) \,
   _5F_4\left(\begin{array}{l}\frac{1}{2}+2
   x,\frac{1}{4}+w+x+y,\frac{1}{2}+2 y,\frac{1}{2}+2
   x+2 y, \frac{1}{2}\cr 1+2 x,\frac{5}{4}+w+x+y,1+2 y,1+2 x+2
   y\end{array};1\right)}{(1+4 w+4 x+4 y) \Gamma (1+2 x) \Gamma
   (1+2 y) \Gamma (1+2 x+2 y)} +
 \cr
   & -\frac{\Gamma (-2 x) \Gamma \left(\frac{1}{2}+2
   x\right) \Gamma (-2 y) \Gamma \left(\frac{1}{2}+2
   y\right) }{32 \pi ^{5/2} (1+4 w+4 x+4 y) \Gamma
   \left(\frac{1}{2}-2 x-2 y\right) \Gamma (1+2 x+2
   y)} \times \cr  &  \times\left(
   \, _6 \ddot F_5\left(\begin{array}{l}\frac{1}{2}+x+y,\frac{1}{2}+2
   x,\frac{1}{4}+w+x+y,\frac{1}{2}+2 y,\frac{1}{2}+2
   x+2 y,\frac{1}{2}\cr \frac{1}{2}+x+y,1+2
   x,\frac{5}{4}+w+x+y,1+2 y,1+2 x+2 y\end{array};1\right) +\right. 
   \cr  &  +\left.
   \, _6 \ddot F_5\left(\begin{array}{l}1+x+y,\frac{1}{2}+2
   x,\frac{1}{4}+w+x+y,\frac{1}{2}+2 y,\frac{1}{2}+2
   x+2 y,\frac{1}{2}\cr 1+x+y,1+2
   x,\frac{5}{4}+w+x+y,1+2 y,1+2 x+2 y\end{array};1\right) \right)\label{222} \cr &
\end{align}
where we introduced the notation
\begin{align}
 {}_p \dot F_q\left(\begin{array}{l} a,\cdots \cr b,\cdots\end{array};z\right) = \frac{ \partial }{\partial a}{}_pF_q\left(\begin{array}{l} a,\cdots \cr b,\cdots\end{array};z\right), \ \ \ \ {}_p \ddot F_q\left(\begin{array}{l} a,\cdots \cr b,\cdots\end{array};z\right) = \frac{ \partial }{\partial b}{}_pF_q\left(\begin{array}{l} a,\cdots \cr b,\cdots\end{array};z\right).
\end{align}
While the above result  is perfectly finite and usable, there should exist an underlying simplification, as  Eqs. (\ref{221})-(\ref{222})  should be  a one-parameter\footnote{The parameter is the de Sitter radius $R$, which in the formula is set to 1} deformation of the flat space formula which may simply be written in terms of dilogarithms \cite{Cacciatori2023a,Vanhove}.
Actually, also in flat space a very natural construction in position space gives a formula containing derivatives of the  hypergeometric function $\ {}_2F_1 \ $  w.r.t. the parameters \cite{Cacciatori2023a}. In that case it is  been possible to go further and proceed to dilogarithms which, by the way, are not more explicit the hypergeometrics. Here the task is much more difficult and at the moment we are happy with Eq. (\ref{222}).

\section{Spacetime dimension  $d=3$}
\subsection*{Flat space recapitulation}
Let us review at first the three-dimensional case  in flat space.
Only the first line in formula (\ref{bf2}) has a  pole at $d=3$, namely  the simple pole in the Gamma function. A Laurent expansion near $d\sim 3$ gives
\begin{align}
& I_3(m_1,m_2,m_3,d)=-\frac{1}{32 \pi ^2 (d-3)}+\frac{2-2 \gamma +\log \left(16 \pi ^2\right)}{64 \pi ^2}+\cr & -\frac{\tanh ^{-1}\left(\frac{2 \a \b}{\a^2+\b^2-\c^2}\right)+\tanh ^{-1}\left(\frac{2 \a
   \c}{\a^2-\b^2+\c^2}\right)+\tanh ^{-1}\left(\frac{2 \b \c}{-\a^2+\b^2+\c^2}\right)}{32 \pi ^2}+\cr &-\frac{\log \left(-\a^4-\b^4-\c^4+2 \a^2 \b^2+2 \a^2 \c^2+2 \b^2 \c^2\right)}{64 \pi ^2} +{\rm O}(d-3).
\end{align}
By using the identity $\tanh ^{-1}(x)=\frac{1}{2} \log \left(\frac{1+x}{1-x}\right)$ the  result simplifies to
\begin{align}
 I_3(m_1,m_2,m_3,d)=&-\frac{1}{32 \pi ^2 (d-3)}-\frac{1}{16 \pi ^2}\log (\a+\b+\c) \cr & 
+\frac{ 1-\gamma+\log (4 \pi )}{32 \pi ^2}+{\rm O}(d-3).\end{align}
Alternatively, we may proceed by  an ultraviolet cutoff of the integral defining the watermelon; in position space we cut a little sphere surrounding the origin $x=0$:
\begin{align}
I_{3,\Lambda}(m_1,m_2,m_3,3)
&= \frac{ \left({\a\b\c}\right)^{\frac{1}{2}} }{ (2\pi)^{\frac{5}{2}} \pi        }
\int_{\Lambda^2}^\infty \  r^{\frac{1}{2}}  K_{\frac{1}{2}}\left( \a r \right)   K_{\frac{1}{2}}\left( \b r \right)    K_{\frac{1}{2}}\left( \c r \right) dr  \label{BB} \cr& =  \int_{\Lambda^2}^\infty\frac{e^{-r (\a+\b+\c)}}{16 \pi ^2 r}dr = \frac{\Gamma (0,(\a+\b+\c) \Lambda^2 )}{16 \pi ^2} .
\end{align}
Expansion in $\Lambda$ now gives
\begin{align}
I_{3,\Lambda}(m_1,m_2,m_3,3)
& = -\frac{\log (\Lambda^4 )}{32 \pi ^2}-\frac 1 {16 \pi ^2}{\log (\a+\b+\c)}-\frac{\gamma }{16 \pi ^2}.
\end{align}
We see that the mass dependent finite parts coincide in the two expression. On the other hand the additive constants depend on the renormalization scheme. 

\subsection*{de Sitter: UV cutoff}

At $d=3$ the Schwinger function is elementary:
\begin{equation}
 G^{3}_{\nu}(- \cos s )  =
\frac{ \sinh (\nu (\pi -s))}{4 \pi \sinh(\pi\nu) \sin s }.
\end{equation}
We may cutoff the  integral as in flat space
\begin{align}
& I_{3,K}(\na,\nb,\nc,d) = 
4\pi  \int _{K}^\pi G^{3}_{\nu_1}(-\cos s )  G^{3}_{\nu_2}(-\cos s )  G^{3}_{\nu_3}(-\cos s )  {\sin^2 s}  {ds} =\cr  & =
{\int_K^\pi 
\frac{\scriptstyle{ \sinh ((\pi -s) (\na -\nb -\nc ))-\sinh
   ((\pi -s) (\na +\nb -\nc ))-\sinh ((\pi -s)
   (\na -\nb +\nc ))+\sinh ((\pi -s) (\na
   +\nb +\nc ))}}{ {64 \pi ^2 \sinh(\pi  \na ) \sinh(\pi  \nb
   ) \sinh(\pi  \nc )}\sin s} ds}. \cr \end{align}
This leads us to consider  the following indefinite integral
\begin{align}
F_a(s)=\int \frac{ \sinh (a (\pi -s)) }{\sin s} {ds} = \frac{1}{2} e^{a s-\pi  a+i s} \Phi \left(e^{2 i
   s},1,\frac{1}{2}-\frac{i a}{2}\right)-\frac{1}{2}
   \psi \left(\frac{1}{2}-\frac{i a}{2}\right)+ \cr  -\frac{1}{2} e^{-a s+\pi  a+i s} \Phi \left(e^{2 i
   s},1,\frac{1}{2}+\frac{i a}{2}\right)+\frac{1}{2}
   \psi \left(\frac{1}{2}+\frac{i a}{2}\right).
\end{align}
Here $\Phi$ is the Lerch transcendent function and  we adjusted the primitive $F_a(s)$ so that at the upper bound it vanishes at $s=\pi$. It follows that 
\begin{align}
\int_{K}^\pi \frac{ \sinh (a (\pi -s)) }{\sin s} {ds} &=- F_a(K)\simeq - \sinh (\pi  a) (\log (2 K)+\gamma )+\cr &-\frac{1}{2} \sinh
   (\pi  a) \left(\psi \left(\frac{1}{2}-\frac{i
   a}{2}\right)+\psi \left(\frac{1}{2}+\frac{i
   a}{2}\right)\right) +{\rm O}(K)
\end{align}
and therefore
\begin{align}
&\int _{K}^\pi G^{3}_{\nu_1}(-\cos s )  G^{3}_{\nu_2}(-\cos s )  G^{3}_{\nu_3}(-\cos s )  {\sin^2 s}  {ds}  
= - \frac{\log (2 K)+\gamma }{16 \pi^2}+
\cr &
-\frac{ 
   \psi \left(\frac{1}{2}+ \frac{i(\na - \nb - \nc)
   }{2}\right)+
      \psi \left(\frac{1}{2}- \frac{i(\na - \nb - \nc)
   }{2}\right)
  }{128 \pi ^2 \sinh(\pi  \na ) \sinh(\pi 
   \nb ) \sinh(\pi  \nc ) }  \sinh (\pi  (\na
   -\nb -\nc ))+\cr 
   & 
   -\frac{ 
   \psi \left(\frac{1}{2}+ \frac{i(\na + \nb + \nc)
   }{2}\right)+
      \psi \left(\frac{1}{2}- \frac{i(\na + \nb + \nc)
   }{2}\right)
  }{128 \pi ^2 \sinh(\pi  \na ) \sinh(\pi 
   \nb ) \sinh(\pi  \nc ) } \sinh (\pi  (\na
   +\nb +\nc ))+\cr &
+\frac{ 
   \psi \left(\frac{1}{2}+ \frac{i(\na - \nb + \nc)
   }{2}\right)+
      \psi \left(\frac{1}{2}- \frac{i(\na - \nb + \nc)
   }{2}\right)
  }{128 \pi ^2 \sinh(\pi  \na ) \sinh(\pi 
   \nb ) \sinh(\pi  \nc ) }  \sinh (\pi  (\na
   -\nb+ \nc ))+\cr 
   & 
   +\frac{ 
   \psi \left(\frac{1}{2}+ \frac{i(\na + \nb - \nc)
   }{2}\right)+
      \psi \left(\frac{1}{2}- \frac{i(\na + \nb - \nc)
   }{2}\right)
  }{128 \pi ^2 \sinh(\pi  \na ) \sinh(\pi 
   \nb ) \sinh(\pi  \nc ) } \sinh (\pi  (\na
   +\nb -\nc ))+{\rm O}(K). \label{regl}
\end{align}
\subsection*{Dimensional regularization}
The task is now to extract the finite part of the watermelon at  $d=3$ from our  general formulae.  We accomplish this task in full detail, as in this example we explain the algorithm to extract information from our general formula for integer dimensions. 
The hypergeometric series at the rhs of Eq. (\ref{ad}) converges for $d<4$ and therefore the first term $I^{(1)}_3$ is finite.
A direct calculation shows that 
\begin{align}
A_{3}( x, y,w )=\frac{\cot (2 \pi  w) (\cot (2 \pi  x)+ \cot (2 \pi  y)) \left(\psi
   \left(\frac{1}{2}+w+x+y\right)-\psi\left(\frac{1}{2}-w+x+y\right)\right)}{128 \pi ^2}.
\end{align}
The full $I_{3}^{(1)}( x, y,w ,3)$ follows from (\ref{i1}). On the other hand 
$I_3^{(2)}$ contains the divergent part of the diagram, here a simple pole:
\begin{align}
&  {I_3^{(2)}}(x,y,w,d )\simeq  \frac{R(x,y,w,3 )}{d-3}+\widetilde {I_3}(x,y,w,3 ). \label{815}
 \end{align}
Both the residue and the finite term may be computed with the help of the contiguity relations explained in Appendix \ref{pole}. 
Details are provided in Appendix \ref{d=3}. In particular the residue of $B_d(x,y,w)$ at $d=3$ is given by
\begin{equation}
 R_B(x,y,w)= 
 b_3(x,y,w) f_3(x,y,w)  =  \frac{1-\cot (2 \pi  x) \cot (2 \pi  y)}{128 \pi ^2} \label{resb}
\end{equation}
and does not depend on $w$ (see Eqs. (\ref{bd}) and (\ref{polef})). In turn, the residue of  $I_3^{(2)}$ at $d=3$ is the sum of four terms:
\begin{align}
 R(x,y,w) =  R_B(x,y,w)+ R_B(x,-y,w)+ R_B(-x,y,w)+ R_B(-x,-y,w)= \frac 1{32 \pi^2} .  \label{resid3}
\end{align}
 We get the same result as in flat space. This fact is indeed nontrivial: the calculation in flat space and  the calculation on the Euclidean sphere  have essentially nothing to do with each other; the latter expresses a global quantity resulting from integrating on the curved sphere  (here the radius of the sphere has been set to 1, but the result does not depend on the curvature) the de Sitter propagators which are completely different from the propagators in flat  space in the large; they are similar to each other only    in a small region as compared with the curvature and for lage value of the masses.

As regards the finite term, calculations are cumbersome; some details may found in Appendix \ref{d=3}. 
 In the special case where $w=0$ (i.e. $\nc=0$),  Eqs. (\ref{c2}), (\ref{polef}), (\ref{c7}) and (\ref{c13}) together give
 \begin{align}
&  \widetilde {B}_3(x,y,0 )  = \cr &=\frac{(\cot (2 \pi  x) \cot (2 \pi  y)-1) \left(1-\gamma -\pi  \cot (4 \pi  (x+y))+\log (\pi )-2 \psi
   \left(\frac{1}{2}+x+y\right)\right)}{128 \pi ^2} \cr & \label{bbdd0}
 \end{align}
which allows for the calculation of $\widetilde {I_3}(x,y,0,3 )$.
We may actually solve for $\widetilde {I_3}(x,y,w,3 )$. in full generality  by observing that the functions 
  \begin{eqnarray}
 &&\Delta I_3^{(2)} =I_3^{(2)}(x,y,w,d) - I_3^{(2)}(x,y,0,d),\\
 &&\Delta B_d=B_d(x,y,w) - B_d(x,y,0),
 \end{eqnarray}
 are regular at $d=3$ because the residues (\ref{resb}) and (\ref{resid3}) do not depend on $w$. In particular, by using Eq. (\ref{general}) and the identity 
 $$  \frac{1}{\left(\kappa ^2-\nu ^2\right)}-
 \frac 1{\kappa^2}=\frac{\nu ^2\Gamma \left(-\frac{i \kappa }{2}\right) \Gamma \left(\frac{i \kappa }{2}\right)}{4 \left(\kappa ^2-\nu ^2\right)
   \Gamma \left(1-\frac{i \kappa }{2}\right) \Gamma \left(\frac{i \kappa }{2}+1\right)}$$
we get
\begin{eqnarray}
\Delta I_3^{(2)} &= &- \frac{ w^2}{2\pi i}\int 
\prod_{\epsilon,\epsilon',\epsilon''=\pm 1}\frac{2^{d-2}\pi \Gamma \left(\epsilon s+w\right)  \Gamma \left(\epsilon s +\delta +x+y\right)}
   {\Gamma \left(1+\epsilon s  \right) \Gamma \left(\epsilon s+\frac{1}{4}\right) \Gamma \left(\epsilon s+\frac{1}{2}\right) \Gamma \left(\epsilon s+\frac{3}{4}\right)  \Gamma \left(\epsilon s+w+1\right)}. \cr & &\label{bubble2}\end{eqnarray}
By expressing 
$\Delta I_3^{(2)}$ in terms of hypergeometrics as in Eq. (\ref{bd}), after some work, we obtain that
 \begin{eqnarray}
  \Delta B_3 = \lim_{d\to 3} (B_d(x,y,w) - B_d(x,y,0))  
  =- \frac{w^2 (\cot (2 \pi x)\cot (2 \pi y)-1) }{8 \pi ^2 (2 x+2 y+1) \left(4 w^2-(2 x+2 y+1)^2\right)}\times \cr  \times {} _4F_3\left(\begin{array}{r} 1,x+y+\frac{1}{2},w+x+y+\frac{1}{2},-w+x+y
   +\frac{1}{2}\cr x+y+\frac{3}{2},w+x+y+\frac{3}{2},-w+x+y+\frac{3}{2}\end{array} ;1\right)  \cr \cr \cr  =
\frac{(\cot (2 \pi x)\cot (2 \pi y)-1) \left(2 \psi \left(\frac{1}{2}+x+y\right)\right)-\psi \left(\frac{1}{2}-w+x+y\right)-\psi
   \left(\frac{1}{2}+w+x+y\right)}{128 \pi ^2 }. \cr \label{bbdd}
 \end{eqnarray}
 Combining Eqs.  (\ref{bbdd0}) and (\ref{bbdd}) we get 
  \begin{align}
&  \widetilde {B}_3(x,y,w )  =\frac{(1-\cot (2 \pi  x) \cot (2 \pi  y)) \left(\pi  \cot (4 \pi  (x+y))-1+\gamma -\log (\pi )\right)}{128 \pi ^2}  \cr &+\frac{(1-\cot (2 \pi  x) \cot (2 \pi  y)) \left(\psi
   \left(\frac{1}{2}-w+x+y\right)+\psi \left(\frac{1}{2}+w+x+y\right)\right)}{128 \pi ^2} 
 \end{align} 
and $\widetilde {I}_2(x,y,w,3 )$ is computed by (\ref{i2}).

 Collecting all the contributions the final result at $d\sim 3$ is given by 
  \begin{align}
&{I}(x,y,w,d\sim 3 ) =- \frac {1}{32 \pi^2 (d-3) } + \frac{1-\gamma +\log (\pi )}{32 \pi ^2} \label{815} + 
\cr & \sum_{\epsilon,\epsilon'=\pm} \frac{\left(\psi\left(\frac{1}{2}-w-\epsilon  x-\epsilon ' y\right)+\psi \left(\frac{1}{2}+w+\epsilon
    x+\epsilon ' y\right)\right) \sin \left(2 \pi  \left(w+\epsilon  x+\epsilon ' y\right)\right)}{128 \pi ^2 \sin
   (2 \pi  w) \sin (2 \pi  \epsilon  x) \sin \left(2 \pi  \epsilon ' y\right)}
 \end{align}
  to be compared with (\ref{regl}): the mass-dependent finite part is the same as in Eq. (\ref{regl}).
  
\section{Spacetime dimension $d=4$
}
In the four dimensional case both 
$I_3^{(1)}$ 
and $I_3^{(2)}$ diverge. The situation is complicated even more  by the presence of a double pole. While the algorithm remains the same,  the whole procedure of extracting the residues and the finite parts from our general formulae through the contiguity scheme is quite heavy and the formulae cumbersome.

We will refrain to reproduce those formulae here. Instead we proceed to identify the residues of the poles but yet another method which has a quite general domain of applicability. We describe the general construction first.

\subsection*{Convergence, poles and residues}
\label{conver}
In Sec. \ref{2loop} we considered integrals of
the form:
\begin{align}
&{\bf J} = \int_{\eta+i\bR} J((a_p), (c_p), s) ds\ ,\cr
& J((a_p), (c_p), s) = {\prod_{j=1}^{p-1}\Gamma(a_j+s)\Gamma(a_j-s)\over
(a_p+s)(a_p-s)\prod_{j=1}^{p-1}\Gamma(c_j+s)\Gamma(c_j-s)} = {\prod_{j=1}^{p}\Gamma(a_j+s)\Gamma(a_j-s)\over
\prod_{j=1}^{p}\Gamma(c_j+s)\Gamma(c_j-s)}\ ,
\label{co.10}\cr & 
\end{align}
where $c_p = a_p+1$. 
 Note that 
$J((a_p), (c_p), s) = J((a_p), (c_p), -s)$.
It is assumed that $\Re a_j >0$ for $1\le j\le p-1$,
$a_p$ is pure imaginary, $\eta >0$ is small, so that all the poles
of ${ J}$ (as a function of $s$) on the right of the contour are those
located at $s= a_j+n$, $n\geq 0$ integer, $1\le j\le p-1$.
We also assume that outside of $\{|s| \le K\}$ the contour is deformed so
that it lies on the imaginary axis $i\bR$. Let 
\beq
\Z = 2\sum_{j=1}^p(a_j-c_j) \ .
\label{co.11}\endq
Since $|\arg s| \le {\pi\over 2}$  we may use the 
Erd\'elyi-Tricomi Theorem
 \cite[Eq. 5.11.13]{NIST}, \cite[pp. 118 ff]{Olver} :
as $s$ tends to  $ \pm i\infty$
\beq
{\Gamma(a_j+s)\over \Gamma(c_j+s)} \sim
s^{a_j-c_j}\sum_{k=0}^\infty G_k(a_j,\ c_j)\,s^{-k}\ 
\label{co.20}\endq
where the first three terms are given by
\begin{align}
&G_0 = 1,\ \ G_1 (a,b)= \half(a-b)(a+b-1) ,\cr
& G_2(a,b) = {1\over 12} \left ( {a-b \atop 2}\right )
\Big ( 3(a+b-1)^2-(a-b+1)\Big )\ .
\label{u.5}\end{align}
In general, $G_n(a,b)$ is a polynomial with real rational coefficients in $a$ and $b$.

It follows that for a fixed integer $N \ge 1$, there is a $L > 0$
and a bounded analytic function $h_N((a_p),(c_p),s)$ such that, for
$s = it$, $t \ge L$,
\beq
J((a_p), (c_p), s) = t^\Z \Big ( \sum_{k=0}^N {u_k\over t^k} \Big )
+ t^{\Z-N-1}h_N((a_p),(c_p),s)\ .
\label{co.21}\endq
Here $u_0 = 1$ and the other $u_k$ are  polynomial expressions of the
$G_n(a_j,c_j)$, $0\le n \le N$, $1\le j \le p$.

Suppose that $\Re \Z < -1$: integrating $J((a_p), (c_p), s)$
over the whole integration contour
gives 
\beq
2i\int_L^\infty t^\Z \Big ( \sum_{k=0}^N {u_k\over t^k} \Big )\,dt
= \sum_{k=0}^N {-2i u_k L^{\Z -k+1}\over \Z-k+1}\ 
\label{co.25}\endq
plus a bounded analytic function of the parameters.

The result has a  meromorphic continuation for $\Re \Z <N$\ ,
with poles given by (\ref{co.25}).
The residue at $\Z = k-1$ is $-2 i u_k$. In particular the residue at
$\Z = -1$ is $-2i$ and is independent of the parameters.
Also,  $u_{2n+1}=0$, since $J((a_p), (c_p), s)$ is an even
function of $s$.   
For $N=2$ we can write
\begin{equation}
J((a_p), (c_p), s) = 
 t^\Z \left ( \Big (1+s^{-2}\sum_{j=1}^p(2G_2(a_j,c_j)-G_1(a_j,c_j)^2)\Big )
+O(s^{-3})\right ).
\label{co.30}\end{equation}
Since $s^2 = -t^2$, we get
\beq
u_2 = \sum_{j=1}^p(G_1(a_j,c_j)^2-2G_2(a_j,c_j))= \sum_{j=1}^p(f_2(a_j)-f_2(c_j)) ,
\label{co.35}\endq
where
\begin{equation}
f_2(a) =  {a(a-1)(2a-1)\over 6}\ .
\label{co.45}\end{equation}
\subsection*{Summary}
{\bf J} extends to a meromorphic function of the parameters having
poles at
\beq
\Z= -1+2n,\ \ \ n\geq 0 \ \ {\rm integer}.
\label{co.60}\endq
Near $\Z = -1+2n$,
\beq
{\bf J} \sim {-2iu_{2n}\over \z+1-2n}\ .
\label{co.65}\endq

\subsection*{Application at $d=3$ and $d=4$}
Let us consider at first the integral (\ref{bubble1a}). It fits with the definition (\ref{co.10}) with   $\Z = 4\delta-4 = d-5$. 
Here we trade the variable $\Z$ for  the spacetime dimension $d$  and conclude that there are poles
 $d = 4 +2n$. The residue of ${\cal A}$ at $d=4$ is thus simply:
\beq
{\rm residue\ of\ } {\cal A}\ \ {\rm at\ \ } d = 4:\ \ \frac{-2 i u_0}{2i}=-1\ .
\label{co.121}
\endq
Also the integral (\ref{bubble2}) fits  with the definition (\ref{co.10}); 
now $\Z = 8\delta-5 = 2d-7$.
There are poles at $d = 3+n$ with residues (in the variable $d$)
 $- 2^{d-2} i u_{2n}/(2i)= -2^{d-3} u_{2n}$; in particular 
\begin{align}
&\hbox{residue of ${\cal B}$ at $d=3$ : \ \ \ }- u_0= -1 \ ,
\label{co.121.2}\cr
&\hbox{residue of ${\cal B}$ at $d=4$ : \ \ \ } - 2 u_2= 2 w^2-2 x^2-2 y^2+\frac{1}{8} \ .
\end{align}
Multiplying by the right normalizations as in Eqs. (\ref{p09o}) and (\ref{p09oi}) and restoring the mass variables  we obtain the leading singularities as follows:
\begin{align}
&\hbox{ ${I_3}$ at $d=3$ : \ \ \ } -  \frac{1}{32\pi^2(d-3)}, \ 
\label{co.121.2}\\ 
&\hbox{ ${I_3}$ at $d=4$ : \ \ \ }   -\frac{4 \na^2+4 \nb^2+4 \nc^2+3}{512 \pi ^4 (d-4)^2}. \ 
\label{co.121.2}
\end{align}
Once more, we notice that the dominant divergence in $d=3$ is exactly the same as in the flat case. \eqref{co.121.2} reproduces the flat dominant divergence in the limit $R\to\infty$.

\section{Conclusions}

In a recent paper we have shown that calculating loop integrals in position space may be advantageous also in flat Minkowski space from several viewpoints. While we believe that that possibility in flat space deserves attention and is not just a luxury or a mathematical ornament, in curved space performing loop calculation in position space is compulsory. Here we have started this program in de Sitter space. Calculations are significantly more intricate than in flat space but can be performed successfully till the end. In a companion paper we present a study of the same diagrams in anti de Sitter space. We have found that in the AdS case things are a little simpler than in the present de Sitterian study because of the presence of a true spectral condition. The two papers together open a new way for precision calculations of QFT in the presence of a cosmological constant.

\section*{Acknowledgments} U. M. and S. L. C. are grateful to the Institut des Hautes Etudes Scientifiques - Bures-sur-Yvette,  for the generous hospitality and support they received while writing this paper.

\newpage
\appendix 
\begin{appendix}

\section{Details about section \ref{1loop}}
\label{details1}
\subsection{The functions $\P_\nu^\mu$ and $\Q_\nu^\mu$}
These functions are solutions of the Legendre equation
\cite[3.2 (1) p. 121]{bateman}:
\begin{align}
&{d\over dz}(1-z^2)w'(z) + C_{\nu,\mu}(z)w(z) = 0\ ,\cr
& C_{\nu,\mu}(z) = \nu(\nu+1)-\mu^2(1-z^2)^{-1}\ .
\label{d.10}\end{align}
They are called "Legendre functions on the cut'' in \cite{bateman}
or Ferrers functions \cite[14.3.1, 14.3.2]{NIST}. They are holomorphic in
the domain
\beq
\Delta_2 = \{z\in \bC\ :\ \Im z \not= 0\ \ {\rm or}\ \ -1<z<1 \}
\label{d.20}\endq
and given there by:
\begin{align}
\P_\beta^\alpha(z) &= {1\over \Gamma(1-\alpha)}
\left({1+z\over 1-z} \right )^{\alpha\over 2}
F \left (-\beta,\ 1+\beta\ ;\ 1-\alpha\ ;\ {1-z\over 2}\right )\ ,
\label{d.30}\\
&= {\Gamma(-\alpha) \over \Gamma(1+\beta-\alpha)\Gamma(-\beta-\alpha)}
\left ({1+z \over 1-z} \right )^{\alpha\over 2}
F \left (-\beta,\ 1+\beta\ ;\ 1+\alpha\ ;\ {1+z\over 2}\right )
\cr
&- \pi^{-1} \sin (\pi\beta)\Gamma(\alpha)
\left ({1-z \over 1+z} \right )^{\alpha\over 2}
F \left (-\beta,\ 1+\beta\ ;\ 1-\alpha\ ;\ {1+z\over 2}\right )\ 
\label{d.40}\\
&=\frac {\tan (\pi  \alpha)+ \tan (\pi \beta )}{\pi }
   \Q^\alpha_\beta(z)+ \frac {\tan (\pi  \alpha)- \tan (\pi \beta )}{\pi }
   \Q^\alpha_{-\beta-1}(z),
\label{d.45}
\\ &
\cr
\Q_\beta^\alpha(z) &= 
{\pi\cos(\pi\alpha)\over 2\sin(\pi\alpha)}\P_\beta^\alpha(z)
-{\pi\Gamma(\beta+\alpha+1)
\over 2\sin(\pi\alpha)\Gamma(\beta-\alpha+1)}\P_\beta^{-\alpha}(z)
\label{d.50}\\
& = \half \Gamma(\alpha) \cos(\alpha \pi)
\left ({1+z\over 1-z}\right )^{\alpha \over 2}
F\left(-\beta,\ \beta+1\ ;\ 1-\alpha\ ;\ {1-z\over 2} \right )\cr
+&{\Gamma(1+\beta+\alpha)\Gamma(-\alpha)\over 2\Gamma(1+\beta-\alpha)}
\left ({1-z\over 1+z}\right )^{\alpha \over 2}
F\left(-\beta,\ \beta+1\ ;\ \alpha+1\ ;\ {1-z\over 2} \right ).
\label{d.55}
\end{align}
(See \cite[3.4 (6) p. 143, 3.2 (15) pp 124-125, 3.4 (13), (10) p. 144]
{bateman}).
From this it follows that:

1) as $z \rightarrow 1$
\begin{align}
\P_\beta^\alpha(z)& \sim {1\over \Gamma(1-\alpha)}
\left ({1-z\over 2}\right )^{-{\alpha\over 2}}\ ,
\label{d.60}\\
\Q_\beta^\alpha(z) &\sim \half \Gamma(\alpha)\cos(\pi\alpha)
\left ({1-z\over 2}\right )^{-{\alpha\over 2}}
\ \ {\rm if}\ \ \Re \alpha > 0\ ,
\label{d.65}
\\
\Q_\beta^\alpha(z) &\sim {\Gamma(1+\beta+\alpha)\Gamma(-\alpha)\over
2\Gamma(1+\beta-\alpha)}
\left ({1-z\over 2}\right )^{{\alpha\over 2}}
\ \ {\rm if}\ \ \Re \alpha < 0\ ;
\label{d.70}\end{align}
2) as $z \rightarrow -1$
\begin{align}
\P_\beta^\alpha(z) &\sim 
-\pi^{-1}\sin(\pi\beta)\Gamma(\alpha)\left ({1+z\over 2}\right )^{-{\alpha\over 2}}
\ \ {\rm if}\ \ \Re \alpha > 0\ \ ,
\label{d.75}\\
\P_\beta^\alpha(z) &\sim 
{\Gamma(-\alpha)\over \Gamma(1+\beta-\alpha)
\Gamma(-\beta-\alpha)}\left ({1+z\over 2}\right )^{{\alpha\over 2}}
\ \ {\rm if}\ \Re \ \alpha < 0\ \ ,
\label{d.80}\\ 
\Q_\beta^\alpha(z) &\sim  -{\Gamma(\alpha) \cos(\pi\beta)\over 2}
\left ({1+z\over 2}\right )^{-{\alpha\over 2}}
\ \ {\rm if}\ \ \Re \alpha > 0\ ,
\label{d.85}\end{align}
\begin{align}
&\Q_\beta^\alpha(z) \sim
{\Gamma(-\alpha)\Gamma(1+\beta+\alpha)\over 
2\sin(\pi\alpha)\Gamma(1+\beta-\alpha)}
\Big [ \sin(\pi\beta) - \cos(\pi\alpha)\sin \pi(\alpha+\beta) \Big ]
\left ({1+z\over 2}\right )^{{\alpha\over 2}} \cr
& = -{\Gamma(-\alpha)\Gamma(1+\beta+\alpha)\cos \pi(\alpha+\beta)
\over 2\Gamma(1+\beta-\alpha)}
\left ({1+z\over 2}\right )^{{\alpha\over 2}}
\ \ {\rm if}\ \ \Re \alpha < 0\ .
\label{d.90}\end{align}

\subsection{Derivation of the main Wronskian equation (\ref{a.23})}
Let $u_1$, $u_2$ be solutions of 
\beq
{d\over dz}(1-z^2)u'_j(z) + B_j(z)u_j(z) = 0\ ,\ \ j=1,\ 2,
\label{c.16}\endq
where $B_j$ is analytic in the domain we consider.
Let $D(z)$ be the determinant
\beq
D(z) = \left | \begin{array}{cc}
u_1(z) & u_2(z)\\
(1-z^2)u'_1(z) & (1-z^2)u'_2(z)\\
\end{array} \right | \ .
\label{c.17}\endq
Then
\beq
{d\over dz}D(z) = \left | \begin{array}{cc}
u_1(z) & u_2(z)\\
-B_1(z)u_1(z) & -B_2(z)u_2(z)\\
\end{array} \right |  = u_1(z)u_2(z)\Big [B_1(z)-B_2(z) \Big ] \ .
\label{c.18}\endq
Thus
\begin{align}
\int_a^b u_1(z)u_2(z)&\Big [ B_1(z)-B_2(z) \Big ] dz =
\Big [ D(z) \Big ]_a^b  = \cr
& = \Big [ (1-z^2)[u_1(z)u'_2(z)-u_2(z)u'_1(z)] \Big ]_a^b\ .
\label{c.19}\end{align}
We now take $u_1 = u_\nu^\mu$, $u_2 = v_\sigma^\rho$ with the understanding that
$u_\nu^\mu$ stands for $\P_\nu^\mu$ or $\Q_\nu^\mu$, and independently
$v_\sigma^\rho$ stands for $\P_\sigma^\rho$ or $\Q_\sigma^\rho$.
Thus $B_1(z) = C_{\nu,\mu}(z)$ and $B_2(z) = C_{\sigma,\rho}(z)$,
and
\beq
B_1(z)-B_2(z) = (\nu-\sigma)(\nu+\sigma+1)+(\rho^2-\mu^2)(1-z^2)^{-1}\ .
\label{c.19.1}\endq
We recall the formulae \cite[3.8 (19),(15) p.161]{bateman}
\beq
(1-z^2){d\P_\nu^\mu(z)\over dz} =
-\nu z\P_\nu^\mu(z) + (\nu+\mu)\P_{\nu-1}^\mu(z)\ ,
\label{c.20}\endq
\beq
\P_{\nu-1}^\mu(z) =
z\P_\nu^\mu(z) +(\nu-\mu+1)(1-z^2)^\half\P_\nu^{\mu-1}(z)\ .
\label{c.21}\endq
Hence
\beq
(1-z^2){d\P_\nu^\mu(z)\over dz} =
\mu z\P_\nu^\mu(z) +(\nu+\mu)(\nu-\mu+1)(1-z^2)^\half\P_\nu^{\mu-1}(z)]\ .
\label{c.22}\endq
As stated in \cite{bateman} the formulae (\ref{c.20}-\ref{c.22})
remain valid when $\P$ is replaced by $\Q$. Therefore, setting $\rho = \mu$,
we get:
\begin{align}
\int_a^b u_\nu^\mu(z)v_\sigma^\mu(z)&(\nu-\sigma)(\sigma+\nu+1)dz =\cr
&=\Big [ (1-z^2)^\half(\sigma+\mu)(\sigma-\mu+1)u_\nu^\mu(z)v_\sigma^{\mu-1}(z)\cr
&-(1-z^2)^\half(\nu+\mu)(\nu-\mu+1)u_\nu^{\mu-1}(z)v_\sigma^\mu(z) \Big ]_a^b\ .
\label{c.23}\end{align}
This is eq. (\ref{a.23}). Recall that here the integration is over an arc
(with extremities $a$ and $b$) contained in the domain
$\Delta_2 = \{z \in \bC\ :\ \Im z \not= 0\ \ {\rm or}\ \ |z| <1\}$,
and that $z\mapsto (1-z^2)^\half$ is the function holomorphic in this
domain and equal to $|1-z^2|^\half$ when $z\in (-1,\ 1)$.

\subsubsection{Case of the functions $P_\nu^\mu$ and $Q_\nu^\mu$}
\label{caseroundPQ}
Some details have to be modified if we now suppose that
$u_\nu^\mu$ stands for $P_\nu^\mu$ or $Q_\nu^\mu$, and independently
$v_\sigma^\rho$ stands for $P_\sigma^\rho$ or $Q_\sigma^\rho$. These
functions\footnote{We use the notations of \cite[3.2 (3),(5) p. 122]{bateman}.
Note that $Q_\nu^\mu$ is not defined where $\Gamma(\nu+\mu+1)$ has a pole.}
will be considered as holomorphic in the domain
$\Delta_1 = \{z \in \bC\ :\ \Im z \not= 0\ \ {\rm or}\ \ z >1 \}$.
Then eqs (\ref{c.20}-\ref{c.22}) have to be replaced by
\beq
(1-z^2){dP_\nu^\mu(z)\over dz} =
-\nu zP_\nu^\mu(z) + (\nu+\mu)P_{\nu-1}^\mu(z)\ ,
\label{c.30}\endq
\beq
P_{\nu-1}^\mu(z) =
zP_\nu^\mu(z) -(\nu-\mu+1)(z^2-1)^\half P_\nu^{\mu-1}(z)\ ,
\label{c.31}\endq
(see \cite[3.8 (10), (5) p. 161]{bateman})
\beq
(1-z^2){dP_\nu^\mu(z)\over dz} =
\mu zP_\nu^\mu(z) -(\nu+\mu)(\nu-\mu+1)(z^2-1)^\half P_\nu^{\mu-1}(z)]\ .
\label{c.32}\endq
Here $z\mapsto (z^2-1)^\half$ is the function holomorphic in $\Delta_1$
equal to $|z^2-1|^\half$ when $z> 1$. Again as stated in \cite{bateman}
eqs (\ref{c.30}-\ref{c.32}) remain valid if $P$ is replaced by $Q$.
Therefore, setting $\rho = \mu$,
we get from (\ref{c.19}):
\begin{align}
\int_a^b u_\nu^\mu(z)v_\sigma^\mu(z)&(\nu-\sigma)(\sigma+\nu+1)dz =\cr
&=\Big [ -(z^2-1)^\half(\sigma+\mu)(\sigma-\mu+1)u_\nu^\mu(z)v_\sigma^{\mu-1}(z)\cr
&+(z^2-1)^\half(\nu+\mu)(\nu-\mu+1)u_\nu^{\mu-1}(z)v_\sigma^\mu(z) \Big ]_a^b\ .
\label{c.33}\end{align}
Here the integration is over an arc contained in $\Delta_1$.

\subsection{The basic case $u_\nu^\mu = \Q_\nu^\mu$, $v_\sigma^\mu = \Q_\sigma^\mu$}
\label{caseQQ}
This case may be considered as basic since by using (\ref{d.45}) it is possible
to obtain the other cases from it.
The formula (\ref{a.23}) becomes in this case:
\begin{align}
(\nu-\sigma)(\nu+\sigma+1)\int_a^b &\Q_\nu^\mu(z)\Q_\sigma^\mu(z)\,dz = \cr
&=\Big [ (\sigma+\mu)(\sigma-\mu+1)(1-z^2)^\half
\Q_\nu^\mu(z)\Q_\sigma^{\mu-1}(z)\cr
&-(\nu+\mu)(\nu-\mu+1)(1-z^2)^\half \Q_\nu^{\mu-1}(z)\Q_\sigma^\mu(z)
\Big ]_a^b.
\label{w.10}\end{align}
As in sect. \ref{1loop} it is sufficient to evaluate the contribution
of the first term in the rhs of (\ref{w.10}) since the
contribution of the second can be obtained by exchanging $\nu$ and
$\sigma$ and a global change of sign.

We again fix $\mu \in (0,\ 1)$.
Using the formulae (\ref{d.65}) and (\ref{d.70}) we find that
as $z\rightarrow 1$ the first term in the rhs of (\ref{w.10})
tends to 
\beq
{\Gamma(\mu)\Gamma(1-\mu)\Gamma(1+\sigma+\mu)\cos(\pi\mu)\over
2\Gamma(1+\sigma -\mu)}\ .
\label{w.20}\endq
As $z\rightarrow -1$ (with a minus sign due to dealing
with the lower bound $a$) this first term contributes:
\beq
{\Gamma(\mu)\Gamma(1-\mu)\Gamma(1+\sigma+\mu)\cos(\pi\nu)
\cos\pi(\sigma+\mu)\over 2\Gamma(1+\sigma-\mu)}\ .
\label{w.25}\endq
The rhs of (\ref{w.10}) is thus given by
\begin{align}
&{\Gamma(\mu)\Gamma(1-\mu)\Gamma(1+\sigma+\mu)
[\cos(\pi\mu) + \cos(\pi\nu)\cos\pi(\sigma+\mu)]\over
2\Gamma(1+\sigma -\mu)}\cr
&-{\Gamma(\mu)\Gamma(1-\mu)\Gamma(1+\nu+\mu)
[\cos(\pi\mu) + \cos(\pi\sigma)\cos\pi(\nu+\mu)]\over
2\Gamma(1+\nu -\mu)}.
\label{w.30}\end{align}
In this expression let us set $\nu = -\sigma -1$. The first term
becomes
\begin{align}
&{\Gamma(\mu)\Gamma(1-\mu)\Gamma(1+\sigma+\mu)
[\cos(\pi\mu) - \cos(\pi\sigma)\cos\pi(\sigma+\mu)]\over
2\Gamma(1+\sigma -\mu)}\cr
&= {\Gamma(\mu)\Gamma(1-\mu)\Gamma(1+\sigma+\mu)\sin\pi(\sigma+\mu)
\sin(\pi\sigma)\over 2\Gamma(1+\sigma -\mu)}\cr
&= -{\pi\Gamma(\mu)\Gamma(1-\mu)\sin(\pi\sigma)\over
2\Gamma(1+\sigma -\mu)\Gamma(-\sigma-\mu)}\ .
\label{w.35}\end{align}
The second term becomes
\begin{align}
&-{\Gamma(\mu)\Gamma(1-\mu)\Gamma(\mu-\sigma)
[\cos(\pi\mu) -\cos(\pi\sigma)\cos\pi(\mu-\sigma)]\over
2\Gamma(-\sigma-\mu)}\cr
&= {\Gamma(\mu)\Gamma(1-\mu)\Gamma(\mu-\sigma)\sin\pi(\mu-\sigma)
\sin(\pi\sigma)\over2\Gamma(-\sigma-\mu)}\cr
&= {\pi\Gamma(\mu)\Gamma(1-\mu)\sin(\pi\sigma)\over
2\Gamma(1+\sigma -\mu)\Gamma(-\sigma-\mu)}\ .
\label{w.40}\end{align}
Thus the expression (\ref{w.30}) vanishes if we set $\nu = -\sigma -1$.
It also obviously vanishes if we set $\sigma = \nu$.
We obtain
\begin{align}
&\int_{-1}^1 \Q_\nu^\mu(z) \Q_\sigma^\mu(z)\,dz =
\Gamma(\mu)\Gamma(1-\mu) \Gamma(1+\nu+\mu)\Gamma(1+\sigma+\mu)
\funame_{\Q\Q}(\nu,\ \sigma,\ \mu)\ ,
\label{w.100}\\
& \funame_{\Q\Q}(\nu,\ \sigma,\ \mu) = 
{1\over (\nu-\sigma)(\nu+\sigma+1)}\times \cr
&\qquad  \qquad  \Bigg [{
\cos(\pi\mu) + \cos(\pi\nu)\cos\pi(\sigma+\mu)\over
2\Gamma(1+\nu+\mu)\Gamma(1+\sigma -\mu)}
-{\cos(\pi\mu) + \cos(\pi\sigma)\cos\pi(\nu+\mu)\over
2\Gamma(1+\sigma+\mu)\Gamma(1+\nu -\mu)} \Bigg ]\ .
\label{w.105}\end{align}
$\funame_{\Q\Q}(\nu,\ \sigma,\ \mu)$ extends to an entire function of
all its arguments. It is symmetric in $\nu$ and $\sigma$ and vanishes
when $\mu = 0$. The equality (\ref{w.100}) extends by analytic continuation
to values of $\mu$ with $-1<\Re \mu < 1$.

\subsection{The case $u_\nu^\mu = \P_\nu^\mu$, $v_\sigma^\mu = \Q_\sigma^\mu$}
\label{casePQ}
This case may be dealt with by two different  methods: the first is to use
the Wronskian equation as in the two preceding cases. The second is to use
eq. (\ref{d.45}) and the result of subsect. \ref{caseQQ}.
The formula (\ref{a.23}) becomes in this case:
\begin{align}
(\nu-\sigma)(\nu+\sigma+1)\int_a^b &\P_\nu^\mu(z)\Q_\sigma^\mu(z)\,dz = \cr
&=\Big [ (\sigma+\mu)(\sigma-\mu+1)(1-z^2)^\half
\P_\nu^\mu(z)\Q_\sigma^{\mu-1}(z)\cr
&-(\nu+\mu)(\nu-\mu+1)(1-z^2)^\half \P_\nu^{\mu-1}(z)\Q_\sigma^\mu(z)
\Big ]_a^b .
\label{q.10}\end{align}
We fix $\mu\in (0,\ 1)$.
To evaluate the rhs of this formula as $z \rightarrow 1$ we use the formulae
(\ref{d.60}-\ref{d.70}):
as $z \rightarrow 1$ the first term in the rhs of
(\ref{q.10})
\iffalse
behaves like
\begin{align}
&{(\sigma+\mu)(\sigma-\mu+1)\Gamma(\sigma+\mu)\over
2\Gamma(2+\sigma-\mu)}\times\cr
&2\left({1-z\over 2}\right)^\half \left({1-z\over 2}\right)^{-{\mu\over 2}}
\left({1-z\over 2}\right)^{{\mu\over 2}-\half}\cr
& \rightarrow {\Gamma(1+\sigma+\mu)\over \Gamma(1+\sigma-\mu)}\ .
\label{q.25}\end{align}
\else
tends to
\beq
{\Gamma(1+\sigma+\mu)\over \Gamma(1+\sigma-\mu)}\ .
\label{q.25}\endq
\fi
The second term in the rhs of (\ref{q.10}) behaves like
${\rm Const.\ } (1-z)^{1-\mu}$ and tends to 0.
To evaluate the rhs of the formula (\ref{q.10}) as $z\rightarrow -1$
we use the formula (\ref{d.75}-\ref{d.90}).
In the end we find that, for $a= -1$ and $b=1$ the rhs of (\ref{q.10})
is equal to
\begin{align}
&{\Gamma(1+\sigma+\mu)\over \Gamma(1+\sigma-\mu)}
\cr
&+{\Gamma(\mu)\Gamma(1-\mu)\Gamma(1+\sigma+\mu)\sin(\pi\nu)\cos\pi(\mu+\sigma)
\over \pi\Gamma(1+\sigma-\mu)}
\cr
&+{\Gamma(\mu)\Gamma(1-\mu)\cos(\pi\sigma)\over
\Gamma(1+\nu-\mu)\Gamma(-\nu-\mu)}\ .
\label{q.80}\end{align}
If we set $\nu = \sigma$ in this expression, it becomes
\begin{align}
&{\Gamma(1+\sigma+\mu)\over \Gamma(1+\sigma-\mu)}
\cr
&-{\Gamma(1+\sigma+\mu) \Gamma(\mu)\Gamma(1-\mu)\over \pi\Gamma(1+\sigma-\mu)}
\times \cr
& \Bigg [ -\sin(\pi\sigma)\cos\pi(\mu+\sigma)
-{\pi\cos(\pi\sigma) \over \Gamma(1+\sigma+\mu)\Gamma(-\sigma-\mu)}\Bigg ] 
= 0\ ,
\label{q.82}\end{align}
as it must be.
We can therefore subtract from (\ref{q.80}) the same expression with
$\nu=\sigma$ and we obtain, for $0< \mu < 1$,
\begin{align}
&\int_{-1}^1 \P_\nu^\mu(z) \Q_\sigma^\mu(z)\,dz =
\Gamma(\mu)\Gamma(1-\mu) \Gamma(1+\sigma+\mu)\funame_{\P\Q}(\nu,\ \sigma,\ \mu)\ ,
\label{q.100}\\
& \funame_{\P\Q}(\nu,\ \sigma,\ \mu) = 
{1\over (\nu-\sigma)(\nu+\sigma+1)}\times \cr
&\Bigg [
{[\sin(\pi\nu)-\sin(\pi\sigma)]
\cos\pi(\mu+\sigma)
\over \pi\Gamma(1+\sigma-\mu)} \cr
&+{\cos(\pi\sigma) \over \Gamma(1+\sigma+\mu)}\left [ {1\over
\Gamma(1+\nu-\mu)\Gamma(-\nu-\mu)} - {1\over
\Gamma(1+\sigma-\mu)\Gamma(-\sigma-\mu)} \right ]
\Bigg ]\ .
\label{q.101}\end{align}
The function $\funame_{\P\Q}(\nu,\ \sigma,\ \mu)$ is entire in all its variables.
It vanishes at $\mu = 0$ and it is invariant under
the change $\nu \rightarrow -1-\nu$. The equality (\ref{q.100}) extends
by analytic continuation to $-1< \Re \mu < 1$.

The second method is to use (\ref{d.45}) to write
\begin{align}
&\int_{-1}^1 \P_\nu^\mu(z) \Q_\sigma^\mu(z)\,dz =\cr &
\frac {\tan (\pi  \mu)+ \tan (\pi \nu )}{\pi }
 \int_{-1}^1 \Q_\nu^\mu(z) \Q_\sigma^\mu(z)\,dz  + \frac {\tan (\pi  \mu)- \tan (\pi \nu )}{\pi }
  \int_{-1}^1 \Q^\mu_{-\nu-1}(z) \Q_\sigma^\mu(z)\,dz =\cr 
& = \frac 1{(\nu -\sigma ) (\nu +\sigma +1)} \left( \frac{\Gamma (\mu +\sigma +1) (\sec (\pi  \mu )
   \cos (\pi  \nu ) \cos (\pi  (\mu +\sigma
   ))+1)}{
   \Gamma (-\mu +\sigma +1)}\right. +\cr & \left.+\frac{\pi  \nu  \sec (\pi  \mu ) \tan (\pi  \nu
   ) \cos (\pi  \sigma )-\pi  \mu  \csc (\pi 
   \mu ) (\sec (\pi  \nu )+\cos (\pi  \sigma
   ))}{\Gamma (-\mu -\nu +1) \Gamma (-\mu +\nu
   +1)}\right).
\label{q.110}\end{align}

\subsection{Other integrals}

Using the formulae \cite[3.4 (14), (15), (17), (18) p. 144]{bateman}
\beq
\P_\beta^\alpha(-z) =\cos\pi(\beta+\alpha)\P_\beta^\alpha(z)
-{2\sin\pi(\beta+\alpha)\over \pi}\Q_\beta^\alpha(z)\ ,
\label{w.150}\endq 
\beq
\Q_\beta^\alpha(-z) = -\cos\pi(\beta+\alpha)\Q_\beta^\alpha(z)<
-{\sin\pi(\beta+\alpha)\over 2\pi}\P_\beta^\alpha(z)\ ,
\label{w.155}\endq 
\beq
\Gamma(1+\beta+\alpha)\P_\beta^{-\alpha}(z) = \Gamma(1+\beta-\alpha) \Big [
\P_\beta^\alpha(z)\cos(\pi\alpha)-{2\over \pi}\sin(\pi\alpha)\Q_\beta^\alpha(z)\Big ]\ ,
\label{w.160}\endq
\beq
\Gamma(1+\beta+\alpha)\Q_\beta^{-\alpha}(z) = \Gamma(1+\beta-\alpha) \Big [
\Q_\beta^\alpha(z)\cos(\pi\alpha) + {\pi\over 2}\sin(\pi\alpha)\P_\beta^\alpha(z)\Big ]\ ,
\label{w.165}\endq
we can immediately obtain
$\int_{-1}^1 \P_\nu^{\pm \mu}(\pm z)\P_\sigma^{\pm \mu}(\pm z) dz$,
$\int_{-1}^1 \P_\nu^{\pm \mu}(\pm z)\Q_\sigma^{\pm \mu}(\pm z) dz$,\break
$\int_{-1}^1 \Q_\nu^{\pm \mu}(\pm z)\Q_\sigma^{\pm \mu}(\pm z) dz$,
for instance :
\begin{align}
&\int_{-1}^1 \P_\nu^\mu(z) \P_\sigma^\mu(-z)\,dz =
2\pi^{-1}\cos\pi(\sigma+\mu)\Gamma(\mu)\Gamma(1-\mu)
\funame_{\P\P}(\nu,\ \sigma,\ \mu) \cr
&-{2\over \pi}\sin\pi(\sigma+\mu)\Gamma(\mu)\Gamma(1-\mu)\Gamma(1+\sigma+\mu)
\funame_{\P\Q}(\nu,\ \sigma,\ \mu)\ .
\label{w.170}\end{align}

Finally let us use subsubsect. \ref{caseroundPQ} to evaluate the integral
\beq
\int_1^\infty Q_\nu^\mu(z) Q_\sigma^\mu(z) dz\ .
\label{wq.10}\endq
It follows from \cite[3.9.2 (21), (5), (6) pp 163-164]{bateman} that:
\beq
{\rm as}\ \ z \rightarrow +\infty,\ \ Q_\nu^\mu(z) \sim {\rm const.\ }
z^{-\nu-1}\ ;   
\label{wq.20}\endq
\beq
{\rm if\ }\Re \mu >0,\ \ {\rm as}\ \ z \rightarrow 1,\ \ Q_\nu^\mu(z) \sim 
e^{i\pi\mu} 2^{{\mu\over 2}-1}\Gamma(\mu)(z-1)^{-{\mu\over 2}}\ ;   
\label{wq.25}\endq
\beq
{\rm if\ }\Re \mu < 0,\ \ {\rm as}\ \ z \rightarrow 1,\ \ Q_\nu^\mu(z) \sim
{e^{i\pi\mu} 2^{-{\mu\over 2}-1}\Gamma(-\mu)\Gamma(\nu+\mu+1)(z-1)^{{\mu\over 2}}
\over \Gamma(\nu-\mu+1)}\ .   
\label{wq.30}\endq
Note also (\cite[3.2 (2) p. 140]{bateman})
\beq
e^{i\pi\mu}\Gamma(\nu+\mu+1)Q_\nu^{-\mu}(z) =
e^{-i\pi\mu}\Gamma(\nu-\mu+1)Q_\nu^\mu(z)\ .
\label{wq.40}\endq
It follows that the integral (\ref{wq.10}) converges if
$\Re(\nu+\sigma) > -1$ and $|\Re \mu| < 1$. For our evaluation we
suppose $\Re(\nu+\sigma) > -1$ and $0< \mu < 1$. We then set
$u_\nu^\mu(z) = Q_\nu^\mu(z)$ and $v_\sigma^\mu = Q_\sigma^\mu$ in
(\ref{c.33}) and let $b$ tend to infinity and $a$ tend to 1.
The result is that, under our assumptions,
\begin{align}
&\int_1^\infty Q_\nu^\mu(z) Q_\sigma^\mu(z)\, dz = \cr
&{e^{2i\pi\mu}\Gamma(\mu)\Gamma(1-\mu)\over
2(\nu-\sigma)(\sigma+\nu+1)}
\Bigg [ {\Gamma(\nu+\mu+1)\over  \Gamma(\nu-\mu+1)}
-{\Gamma(\sigma+\mu+1)\over  \Gamma(\sigma-\mu+1)} \Bigg ]\ .
\label{wq.50}\end{align}
This equation remains valid, by analytic continuation, when $|\Re \mu| <1$
(and $\Re(\nu+\sigma) > -1$) (note that the bracket vanishes for $\mu=0$).
The formula is compatible with (\ref{wq.40}). Letting $\nu$ tend to
$\sigma$ we find
\begin{align}
\int_1^\infty &Q_\sigma^\mu(z) Q_\sigma^\mu(z)\, dz = \cr
&={e^{2i\pi\mu}\Gamma(\mu)\Gamma(1-\mu)\Gamma(\sigma+\mu+1)\over
2(2\sigma+1)\Gamma(\sigma-\mu+1)}\Big [\psi(\sigma+\mu+1)-\psi(\sigma-\mu+1)
\Big ]\ .
\label{wq.60}\end{align}
Again this equation is valid when $|\Re \mu| <1$ and $\Re 2\sigma > -1$,
but its r.h.s. can be continued outside of this region.


\section{The constant $D$}\label{D}
We have
\begin{align}
    D=& B(0)= \int_0^\infty \frac {2t^4+t^2-1+}{t^4+t^2} \frac {t^3 dt}{e^{2\pi t}-1} \cr
    &+6\int_0^\infty \arctan \left(t\right) \frac {t^2 dt}{e^{2\pi t}-1}\cr
    &+2\int_0^\infty \left(t^2-\frac 12\right) \log [t^2+t^4] \frac {t dt}{e^{2\pi t}-1}. 
\end{align}
The first line gives
\begin{align}
    \int_0^\infty \Bigg( 2-\frac 1{t^2} \Bigg) \frac {t^3 dt}{e^{2\pi t}-1}=2\zeta(4)\frac {\Gamma(4)}{(2\pi)^4}-\frac {\zeta(2)}{(2\pi)^2}=-\frac 1{30}.
\end{align}
For the second line we use
\begin{align}
    \int_0^\infty \arctan \left(t\right) \frac {t^2 dt}{e^{2\pi t}-1}=-\frac 29+\frac {\zeta(3)}{8\pi^2}+\frac 14 \log (2\pi)-\log A,
\end{align}
where $A$ is the Glaisher-Kinkelin constant related to the first derivative of the Riemann $\zeta$ function by \cite{Wang}
\begin{align}
    \log A=\frac 1{12} -\zeta'(-1).
\end{align}
For the last line we use
\begin{align}
    \int_0^\infty (2t^2-1) \log \left(1+t^2\right) \frac {t^2 dt}{e^{2\pi t}-1}=& \frac {11}8 -\frac 32 \left( \log (2\pi) +\frac {\zeta(3)}{\pi^2} \right)
    -2\zeta'(-3)+5\log A, \\
    \int_0^\infty (2t^2-1) \log t^2  \frac {t^2 dt}{e^{2\pi t}-1}=& \frac 1{60}\left(-\frac {19}{6} +\gamma -\log (2\pi) \right)+\frac {3\zeta'(4)}{2\pi^4}
    +\log A.
\end{align}
The standard functional relation for $\zeta(z)$ gives
\begin{align}
    \zeta'(-3)=-\frac {4!}{2(2\pi)^4} \zeta'(4)+\frac {B_4}4(H_3-\gamma-\log(2\pi)).
\end{align}
On the other hand, we can use the Adamchik formula \cite{Adamchik}
\begin{align}
    \zeta'(-3)=\frac {B_4}4 H_3-\log A_3,
\end{align}
where $A_3$ is the third generalized Glaisher-Kinkelin number (or third Bendersky's number) defined by \cite{Wang}
\begin{align}
    \log A_3=&\lim_{n\to\infty} \Bigg[ \sum_{k=1}^n k^3\log k -\left( \frac {n^4}4 +\frac {n^3}2 +\frac {n^2}4 -\frac 1{120} \right)\log n+\frac {n^4}{16}-\frac {n^2}{12} \Bigg]\cr
    \simeq & -0.02065635,
\end{align}
$H_n$ are the harmonic numbers, and $B_n$ the Bernoulli numbers. Summing up, we get
\begin{align}
    D=-\frac 1{72}+\frac {\gamma-\log(2\pi)}{60}-\frac {3\zeta(3)}{4\pi^2}+4\log(A_3)\approx -0.2088707.
\end{align}

\section{Poles from contiguity in hypergeometrics} \label{pole}
Recall that
\beq
{}_{q+1}F_q \left ( \left .\begin{array}{l}
a_1,\ \dots ,\ a_{q+1}\\
b_1,\ \dots ,\ b_q\\ \end{array} \right | z \right )
= \sum_{n=0}^\infty {(a_1)_n\dots (a_{q+1})_n \, z^n\over
n!\,(b_1)_n\dots (b_q)_n}\ .
\label{a.180}\endq
Define
\beq
Z= a_1 + \cdots +a_{q+1} - b_1-\cdots -b_q\ ,
\label{a.181}\endq
then 
\beq
\begin{array}{llll}
\hbox{the series is} &\hbox{absolutely convergent for}& |z| < 1&\\
&\hbox{absolutely convergent for}& |z| = 1
& {\rm if}\ \ \Re Z < 0\ ,\\
&\hbox{conditionally convergent for}& |z| = 1,\ z\not= 1&
{\rm if}\ \ 0 \le \Re Z < 1\ .
\end{array}
\label{a.182}\endq

A general fact for ${}_{q+1}F_q$ is provided by the following formula
[PBM3 p. 441 (37)](after a small correction), [Rainville (19) p. 719]
\begin{align}
& \left [ \sigma +z\sum_{j=1}^q (a_j-b_j)\right ]
{}_{q+1}F_q\left( \left .\begin{array}{l}
(a_q),\ \sigma\\
(b_q)\\ \end{array} \right | z \right ) +\cr
& +z \sum_{j=1}^q 
{(b_j-\sigma)\prod_{k=1}^q(b_j-a_k)\over
b_j\prod_{k=1\atop k\not= j}^q(b_j-b_k)}
{}_{q+1}F_q\left( \left .\begin{array}{l}
(a_q),\ \sigma\\
b_1,\ldots,b_{j-1},b_j+1,b_{j+1},\ldots, b_q\\ \end{array} \right | z \right )\cr
& = \sigma (1-z)\,{}_{q+1}F_q\left( \left .\begin{array}{l}
(a_q),\ \sigma+1\\
(b_q)\\ \end{array} \right | z \right ).
\label{r.200}\end{align}
This requires $b_j\not= b_k$ for all $j\not= k$.
Denote
\beq
Z = \sigma +\sum_{j=1}^q (a_j-b_j)\ .
\label{r.201}\endq
First take the values of the parameters such that
\begin{align}
&\Re Z < -1\ ,\cr
& b_j -b_k \not= 0\ \ \forall\ \ j\not= k\ ,\ \ -b_j \not\in \bZ_+\ .
\label{r.205}\end{align}
In this case it is legitimate to let $z$ tend to 1 everywhere and we get
\begin{align}
& (Z)\ 
{}_{q+1}F_q\left( \left .\begin{array}{l}
(a_q),\ \sigma\\
(b_q)\\ \end{array} \right | 1 \right ) = \cr
& - \sum_{j=1}^q 
{(b_j-\sigma)\prod_{k=1}^q(b_j-a_k)\over
b_j\prod_{k=1\atop k\not= j}^q(b_j-b_k)}
{}_{q+1}F_q\left( \left .\begin{array}{l}
(a_q),\ \sigma\\
b_1,\ldots,b_{j-1},b_j+1,b_{j+1},\ldots, b_q\\ \end{array} \right | 1 \right )
\ .\cr
\label{r.210}\end{align}
The rhs continues to be regular provided 
\begin{align}
&\Re(1-Z) = \sum_{j=1}^q\Re b_j - \sum_{j=1}^q \Re a_j -\Re \sigma +1 > 0\ ,\cr
& b_j -b_k \not= 0\ \ \forall\ \ j\not= k\ ,\ \ -b_j \not\in \bZ_+\ .
\label{r.215}\end{align}
so that, in this region,
\beq
{}_{q+1}F_q\left( \left .\begin{array}{l}
(a_q),\ \sigma\\
(b_q)\\ \end{array} \right | 1 \right ) = 
{\Phi_0((a_q),\ \sigma,\ (b_q))\over Z}\ ,
\label{r.220}\endq
where $\Phi_0((a_q),\ \sigma,\ (b_q))$ is analytic in (\ref{r.215}).
\vsk
This process can be iterated, i.e. it can be applied to each of the
hypergeometic functions appearing in the rhs of (\ref{r.210}), and it
can be iterated further, so that, for any integer $n \ge 0$,
there is a function $\Phi_n((a_q),\ \sigma,\ (b_q))$,
analytic in
\begin{align}
&\Re(n+1-Z)  > 0\ ,\cr
& b_j -b_k \notin \bZ \ \forall\ \ j\not= k\ ,\ \ -b_j \not\in \bZ_+\ ,
\label{r.225}\end{align}
such that, in this region,
\beq
{}_{q+1}F_q\left( \left .\begin{array}{l}
(a_q),\ \sigma\\
(b_q)\\ \end{array} \right | 1 \right ) = 
{\Phi_n((a_q),\ \sigma,\ (b_q))\over Z(Z-1)\ldots (Z-n)}\ .
\label{r.230}\endq
Of course some of the poles appearing in this formula may not actually
occur, i.e. their residues might be zero.

\begin{remark}\rm
\label{rem3}
Let $f$ be a function meromorphic on a domain $D \subset \bC$ that has
at $a\in D$ a Laurent expansion
\beq
f(z) = {f_{-1}\over z-a} +f_0 + f_1(z-a) + \ldots + f_n(z-a)^n + \ldots
\label{r.280}\endq
Then we have
\beq
f_{-1} = (z-a)f(z)\Big |_{z=a}\ ,\ \ \
f_0 = \Big ({d\over dz}\Big )(z-a)f(z) \Big |_{z=a}\ ,\ \ \
f_1 = \half\Big ({d\over dz}\Big )^2(z-a)f(z) \Big |_{z=a}\ ,\ \ \ {\rm etc.}
\label{r.285}\endq
In case e.g. $f$ is a hypergeometric function that, as a function of
$Z$, as above, has a pole at $Z=a$, the contiguity formula (possibly
iterated) may provide an expression for $(Z-a)f(Z)$, as a sum of
convergent hypergeometric series, allowing a computation of the first
terms of the Laurent series of $f$ at $a$.
If
\beq
f(z) = {f_{-p}\over (z-a)^p} + \dots +
{f_{-1}\over z-a} +f_0 + f_1(z-a) + \ldots + f_n(z-a)^n + \ldots
\label{r.290}\endq
then
\beq
f_n = {1\over (n+p)!}\Big ({d\over dz}\Big )^{n+p}(z-a)^p f(z) \Big |_{z=a}\ ,
\ \ \ n \ge -p\ .
\label{r.300}\endq

\end{remark}

\vskip 5 pt

\newpage

\section{ Details about the derivation of the finite term in the three-dimensional case }
\label{d=3}
The hypergeometric series in (\ref{bd}) converges absolutely for $ 
4 -8\delta = 2(3-d)>0.$
The residue and the finite term at $d=3$ are computed as follows:
 \begin{align}
{R_B}(x,y,w,3 )&= \lim_{d\to 3} \left((d-3)B_d(x,y,w)\right),
\\
 \widetilde {B}_3(x,y,w )&= \lim_{d\to 3} \ \partial_d\left((d-3)B_d(x,y,w)\right)  \cr & = b_3( x, y,w)F_3( x, y,w)
  +f_3(x,y,w) \partial_d b_d( x, y,w) \label{c2}
 \end{align}
where 
\begin{align}
&
f_3(x,y,w) =  \lim_{d\to 3} \ \ (d-3)  \  {} _7F_6\left(\begin{array}{c} 2 \delta ,\ldots ,2 \delta +2  u   \cr 2 x+1,\ldots ,\delta +w+ u+1\end{array} ;1\right), \\ & 
F_3(x,y,w) =\lim_{d\to 3} \ \frac {\partial} {\partial d}  \   (d-3) \, {} _7F_6\left(\begin{array}{c} 2 \delta ,\ldots ,2 \delta +2  u   \cr 2 x+1,\ldots ,\delta +w+ u+1\end{array} ;1\right).
 \end{align}
To compute the above quantities the first step is to  rewrite the hypergeometric functions at the rhs of Eq. (\ref{bd}) by using the contiguity relation 

\begin{align}
  & (d -3) \, _7F_6\left(\begin{array}{l}2 \delta ,1+u+\delta ,u-w+\delta ,u+w+\delta ,2 u+2
   \delta ,2 x+2 \delta ,2 y+2 \delta \cr 1+2 x ,1+2 y,1+2 u 
   ,u+\delta  ,1+u-w+\delta  ,1+u+w+\delta \end{array} ;1\right) =
   \cr &=
  \frac{\scriptstyle(2 \delta -1) (1+2 x-2 \delta ) (1-2 y-2 \delta ) (1+2 x-2 y-2 \delta )
   (x-y-\delta ) (1-w+x-y-\delta ) (1+w+x-y-\delta ) \, _7F_6(2 \delta ,\ldots ;2+2
   x,\ldots 
   ;1)}{8 (1+2 x) (x-y) y (1+x-y-\delta ) (w+x-y-\delta  ) (w-x+y+\delta
    )}
    \cr &
    + \frac{\scriptstyle(2 \delta -1) (1+2 y-2 \delta ) (x-y+\delta ) (1+w-x+y-\delta )
   (1-w-x+y-\delta ) (1-2 x-2 \delta ) (1-2 x+2 y-2 \delta ) \, _7F_6(2 \delta
   ,\ldots ;2+2 y,\ldots ;1)}{8 x (x-y) (1+2 y) (1-x+y-\delta ) (w-x+y-\delta
    ) (w+x-y+\delta  )}
   \cr&
  -
  \frac{\scriptstyle(2 \delta -1) (1+2 x-2 \delta ) (1+2 y-2 \delta ) (1+2 x+2 y-2 \delta )
   (x+y-\delta ) (1-w+x+y-\delta ) (1+w+x+y-\delta ) \, _7F_6(2 \delta ,\ldots;2+2 x+2 y,\ldots
   ;1)}{8 x y (1+2 x+2 y) (1+x+y-\delta ) (w-x-y+\delta ) (w+x+y-\delta )}
   \cr& 
   +\frac{\scriptstyle w^2 (x-y-\delta ) (x+y-\delta ) (x-y+\delta ) \, _7F_6\left( 2 \delta ,\ldots;1+x+y+\delta,\ldots;1\right)}{2 (1+x-y-\delta )
   (1-x+y-\delta ) (1+x+y-\delta ) (1-w) (1+w )} \cr &
   -\frac{\scriptstyle(1+2 w) (1+w-x-y-\delta ) (1+w+x-y-\delta ) (1+w-x+y-\delta ) (1+w+x+y-\delta
   ) \, _7F_6(2 \delta ,\ldots ;2+w+x+y+\delta,\ldots ;1)}{ 4 (1+w) (w-x-y+\delta )
   (w+x-y+\delta ) (w-x+y+\delta ) (1+w+x+y+\delta )}
   \cr& +
   \frac{\scriptstyle(1-2 w) (1-w-x-y-\delta ) (1-w+x-y-\delta ) (1-w-x+y-\delta ) (1-w+x+y-\delta
   ) \, _7F_6(2 \delta ,\ldots ;2-w+x+y+\delta
   ,\ldots ;1)}{ 4 (1-w) (w+x+y-\delta ) (w-x+y-\delta )
   (w+x-y-\delta ) (1-w+x+y+\delta )}.
  \label{rtrt}
\end{align}
At the r.h.s., we have only specified the parameters of the hypergeometric functions which are shifted by 1;  the hypergeometric functions at the r.h.s. converge for $d<7/2$. Eq. (\ref{rtrt}) gives us immediately the residue at $d=3$ of the hypergeometric function at the l.h.s.:
\begin{align}& f_3(x,y,w) 
 =\frac{4 w^2-(2 x+2 y+1)^2}{8 x+8 y+4}\label{polef}.
\end{align}
Calculating the finite part is more difficult. Let us start with the easiest:
\begin{eqnarray}
&& \left. f_3(x,y,w) \,  \partial_d b_d( x, y,w)\right|_{d= 3} =  -\frac{(2 x+2 y+1) (\cot (2 \pi  x) \cot (2 \pi  y)-1)}{128 \pi ^2 \left(4 w^2-(2 x+2 y+1)^2\right)} +\cr&&  -\frac{(\cot (2 \pi  x) \cot (2 \pi  y)-1) \left(\psi (2 x+1)+\psi (2 y+1)-\gamma +2+\log
   \left(4 \pi ^2\right)\right)}{256 \pi ^2}+\cr&& +\frac{(\cot (2 \pi  x) \cot (2 \pi  y)-1) \left(\psi \left(2 x+2 y+\frac{3}{2}\right)-2 \psi
  (-4 x-4 y-2)\right)}{256 \pi ^2}. \label{c7}
\end{eqnarray}
On the other hand computing $F_3(x,y,w)$ is quite involved.
 The first three lines  at the r.h.s. of  (\ref{rtrt}) may be grouped into  a single expression that we denote $(d-3 )R_{123}$; their contribution to $F_3(x,y,z)$  is therefore
 \begin{align}
 &R_{123}(x,y,w) = 
 \cr& 
=\frac{\scriptscriptstyle x (2 x-2 y-1) \left((1+2 x+2 y)^2-4 w^2\right) \left((2+4 x-4 y) \psi
   (1+2 x)-(1-2 w+2 x-2 y) \psi \left(\frac{1}{2}-w+x+y\right)-(1+2 w+2
   x-2 y) \psi \left(\frac{1}{2}+w+x+y\right)\right)}{4 \left((1+2 x)^2-4
   y^2\right) \left((1-2 x+2 y)^2-4 w^2\right)} +
   \cr&
  \frac{\scriptscriptstyle y (2 y-2 x-1) \left((1+2 x+2 y)^2-4 w^2\right) \left((2-4 x+4 y) \psi
  (1+2 y)-(1-2 w-2 x+2 y) \psi \left(\frac{1}{2}-w+x+y\right)-(1+2 w-2
   x+2 y) \psi \left(\frac{1}{2}+w+x+y\right)\right)}{4 \left((1+2 y)^2-4
   x^2\right) \left((1+2 x-2 y)^2-4 w^2\right)}+
   \cr& \frac{\scriptscriptstyle(x+y) (2 x+2 y-1) \left((1+2 x+2 y)^2-4 w^2\right) \left((1-2 w+2 x+2 y) \psi
   \left(\frac{1}{2}-w+x+y\right)+(1+2 w+2 x+2 y) \psi
   \left(\frac{1}{2}+w+x+y\right)-2 (1+2 x+2 y) \psi (1+2 x+2
   y)\right)}{4 (1+2 x+2 y)^2 \left(4 w^2-(1-2 x-2 y)^2\right)}.
    \cr&
\end{align}
The other terms are trickier as they involve derivatives of the hypergeometric functions. 
Let us consider the fourth line. After some work we get the following expression:
\begin{align}
 & R_4=\frac{\scriptstyle w \left((1-2 x-2 y) \left((1+2 x+2 y)^2-4 w^2\right)\right)
   \left(\left(H_{-\frac{1}{2}-w+x+y}\right){}^2-\left(H_{-\frac{1}{2}+w+x+y}\right){}^2\right)}{
   32 \left(w^2-1\right) (1+2 x+2 y)}\cr & 
  + \frac{\scriptstyle w \left(2 w^2 \left(-3+12 x^2+8 x y+12 y^2\right)+(1+2 x+2 y)^2 \left(2-x-8 x^2+4 x^3-y-4
   x^2 y-8 y^2-4 x y^2+4 y^3\right)\right) \psi \left(\frac{1}{2}+w+x+y\right)}{4
   \left(-1+w^2\right) (1+2 x+2 y)^2 \left(-1+4 x^2-8 x y+4 y^2\right)}
   \cr &-\frac{\scriptstyle w \left(2 w^2 \left(-3+12 x^2+8 x y+12 y^2\right)+(1+2 x+2 y)^2 \left(2+4 x^3-y-8 y^2+4
   y^3-4 x^2 (2+y)-x \left(1+4 y^2\right)\right)\right) \psi
   \left(\frac{1}{2}-w+x+y\right)}{4 \left(-1+w^2\right) (1+2 x+2 y)^2 \left(-1+4 x^2-8 x
   y+4 y^2\right)}
   \cr &
    +\frac{\scriptstyle 2 w^2 \left(-\frac{1}{2}+x+y\right) \left(
   {}_4\dot F_3\left(\left\{1+2 x,1,\frac{1}{2}-w+x+y,
   \frac{1}{2}+w+x+y\right\},\left\{1+2
   x,\frac{3}{2}-w+x+y,\frac{3}{2}+w+x+y\right\},1\right)\right)}{4 (-1+w) (1+w)
   \left(\frac{1}{2}+x+y\right)}\cr & +\frac{\scriptstyle 2 w^2 \left(-\frac{1}{2}+x+y\right) \left(
   {}_4\dot F_3\left(\left\{1+2 y,1,\frac{1}{2}-w+x+y,
   \frac{1}{2}+w+x+y\right\},\left\{1+2
   y,\frac{3}{2}-w+x+y,\frac{3}{2}+w+x+y\right\},1\right)\right)}{4 (-1+w) (1+w)
   \left(\frac{1}{2}+x+y\right)}\cr & +\frac{\scriptstyle 2 w^2 \left(-\frac{1}{2}+x+y\right) \left(
   {}_4\dot F_3\left(\left\{1+2 x+2y,1,\frac{1}{2}-w+x+y,
   \frac{1}{2}+w+x+y\right\},\left\{1+2
   x+2y,\frac{3}{2}-w+x+y,\frac{3}{2}+w+x+y\right\},1\right)\right)}{4 (-1+w) (1+w)
   \left(\frac{1}{2}+x+y\right)}. \cr& 
\end{align}
As regards the fifth and the sixth line it is useful to compute at first the following auxiliary expression
\begin{align}
    \,
&   _4F_3\left(\begin{array}{r
}s,\frac{3}{2}+u,\frac{1}{2}+u-w,\frac{1}{2
   }+u+w\cr \frac{1}{2}+u,\frac{3}{2}+u-w,\frac{5}{2}+u+w\end{array};1
   \right)=  \cr &= \frac{\Gamma (1-s) \left(\frac{(1+2 u+2 w) (3+2 u+2 w)
   \Gamma \left(\frac{3}{2}+u-w\right)}{\Gamma
   \left(\frac{3}{2}-s+u-w\right)}-\frac{(1+2 u-2 w)
   (-1+2 u-2 w+s (2+4 w)) \Gamma
   \left(\frac{5}{2}+u+w\right)}{\Gamma
   \left(\frac{5}{2}-s+u+w\right)}\right)}{4 (1+2 u)
   (1+2 w)} \cr && 
\end{align}
which allows to compute   $ _4F_3\left(\begin{array}{r
}1,\frac{3}{2}+u,\frac{1}{2}+u-w,\frac{1}{2
   }+u+w\cr \frac{1}{2}+u,\frac{3}{2}+u-w,\frac{5}{2}+u+w\end{array};1
   \right)$, its derivative w.r.t $u$ and 
 $ _4\dot F_3\left(\begin{array}{r
}1,\frac{3}{2}+u,\frac{1}{2}+u-w,\frac{1}{2
   }+u+w\cr \frac{1}{2}+u,\frac{3}{2}+u-w,\frac{5}{2}+u+w\end{array};1
   \right)$.
  \\
We get in this way  an expression for $R_5$ where, as before, there remain  three terms expressed as derivatives of hypergeometric functions: 

\begin{align}
&R_5(x,y,w) =\frac{\scriptstyle(1+2 w+2 x+2 y) \left(4 w^2+4 w^3+(1+2 x+2 y)^2 (w+2 (1+x+y))\right) }{32 (1+w) (1+2 x+2 y)^2
   (3+2 w+2 x+2 y)} 
     (H_{-\frac{1}{2}-w+x+y}-H_{\frac{1}{2}+w+x+y})
     \cr& +\frac{\gamma  \left(4 w^2-(1+2 x+2 y)^2\right) \left(2+4 w+(1+2 w+2 x+2 y)
   \left(H_{\frac{1}{2}+w+x+y}-H_{-\frac{1}{2}-w+x+y}\right)\right)}{64 (1+w) (1+2 x+2
   y)} \cr&+ \frac{(1-2 w+2 x+2 y) (1+2 w+2 x+2 y)^2 \left(\psi
   \left(\frac{1}{2}-w+x+y\right)^2-\psi
  \left(\frac{3}{2}+w+x+y\right)^2\right)}{128 (1+w) (1+2 x+2 y)} +
  \cr&+\frac{\scriptstyle\left(5+8 w-\frac{8 (1+w) (1+2 w)}{1+2 w-2 x-2 y}-\frac{8 (1+w+x) (1+2 w+2 x)}{1+2
   w+2 x-2 y}-\frac{8 x (1+2 x)}{1+2 w-2 x+2 y}\right) 
  }{16 (1+w) (1+2 x+2 y) (6+4 w+4 x+4 y)}\left((1+2 x+2 y)^2-4 w^2\right)\times  \cr & \times  \left(\psi \left(\frac{1}{2}-w+x+y\right)+(1+2 w) \psi
\left(\frac{1}{2}+w+x+y\right)-2 (1+w) \psi
   \left(\frac{3}{2}+w+x+y\right)\right) \cr&
   +\frac{\scriptscriptstyle(1+2 w) (1+2 w+2 x+2 y) \left(4 w (1+w)+(1+2 x+2 y)^2-(1-2 w+2 x+2 y) (1+2 x+2 y)
   (3+2 w+2 x+2 y) \left(H_{\frac{1}{2}+w+x+y}-\gamma \right)\right)}{32 (1+w) (1+2 x+2
   y)^2 (3+2 w+2 x+2 y)} \cr & -\frac{\scriptscriptstyle(1+2 w) (1+2 w+2 x+2 y)\ 
 {}_5 \dot F_4\left(\left\{1+2
   x,\frac{3}{2}+x+y,\frac{1}{2}-w+x+y,\frac{1}{2}+w+x+y,1\right\},\left\{1+2
   x,\frac{1}{2}+x+y,\frac{3}{2}-w+x+y,\frac{5}{2}+w+x+y\right\},1\right)}{8 (1+w) (3+2
   w+2 x+2 y)} \cr &-\frac{\scriptstyle(1+2 w) (1+2 w+2 x+2 y)\ \left(
 {}_5 \dot F_4\left(\left\{1+2
   y,\ldots\right\},\left\{1+2
   y,\ldots\right\},1\right)+ {}_5 \dot F_4\left(\left\{1+2x+2y
   ,\ldots\right\},\left\{1+2x+2
   y,\ldots\right\},1\right)\right)}{8 (1+w) (3+2
   w+2 x+2 y)} .
\end{align}
Finally, 
$R_6(x,y,w)=R_5(x,y,-w)$.
Collecting all the above terms we are able to fully construct $F_3(x,y,w)$ determine $B_3(x,y,w)$ and finally $\tilde{I_2}(x,y,w)$. The result is however utterly complicated. It simplifies a lot when $w=0$ by taking into account the following identity:
\begin{align}
{}_3\dot F_2\left(\begin{array}{l}a,b ,1\cr a,b+2\end{array}; 1\right) = \left.\frac d {ds}  F_2\left(\begin{array}{l}a+s ,b   ,1\cr a,b+2\end{array}; 1\right)\right|_{s=0} = \sum_{n=1}^\infty
\frac{b (1+b) (\psi (a+n)-\psi (a))}{(b+n) (1+b+n)} \cr  = \sum _{n=1}^{\infty } \sum _{k=0}^{n-1} \frac{b (1+b)}{(b+n) (1+b+n) (k+a)} = \sum _{k=0}^{\infty} \sum _{n=k+1}^{\infty } \frac{b (1+b)}{(b+n) (1+b+n) (k+a)} \cr = \sum _{k=0}^{\infty } \frac{b (1+b)}{(a+k) (1+b+k)} = \frac{b (1+b) (\psi(1+b)-\psi (a))}{1-a+b}.
 \end{align}
Collecting everything we land on an indeed very simple result:

\begin{align}
& F(x,y,0) = \frac{(2 x+2 y+1) }{8} \left(H_{2 x+2y}+H_{2 x}+H_{2
   y}-4
   H_{x+y-\frac{1}{2}}\right)-\frac 18.
\label{c13} \end{align}

\end{appendix}

\end{document}